\documentclass[sigplan]{acmart}

\copyrightyear{2024}
\acmYear{2024}
\setcopyright{rightsretained}
\acmConference[SOSP '24]{ACM SIGOPS 30th Symposium on Operating Systems Principles}{November 4--6, 2024}{Austin, TX, USA}
\acmBooktitle{ACM SIGOPS 30th Symposium on Operating Systems Principles (SOSP '24), November 4--6, 2024, Austin, TX, USA}
\acmDOI{10.1145/3694715.3695975}
\acmISBN{979-8-4007-1251-7/24/11}

\usepackage{prp-macros}
\usepackage{mawa-macros}

\newcommand{\sys}{\textsc{Tenplex}\xspace}

\newcommand{\mdp}{multi-dimensional parallelism\xspace}

\newcommand{\ptc}{\mathop{\mathrm{PTC}}}
\newcommand{\ptcc}{$\mathrm{PTC}$\xspace}
\newcommand{\ptcn}{$\mathrm{PTC'}$\xspace}
\newcommand{\tde}{\textsf{Torch Distributed Elastic}}

\begin{document}

\title{\sys: Dynamic Parallelism for Deep Learning\\ using Parallelizable Tensor Collections}


\author{Marcel Wagenländer}
\affiliation{%
  \institution{Imperial College London}
  \country{}
}

\author{Guo Li}
\affiliation{%
  \institution{Imperial College London}
  \country{}
}

\author{Bo Zhao}
\affiliation{%
 \institution{Aalto University}
 \country{}
}

\author{Luo Mai}
\affiliation{%
 \institution{University of Edinburgh}
 \country{}
}

\author{Peter Pietzuch}
\affiliation{%
 \institution{Imperial College London}
 \country{}
}

\renewcommand{\shortauthors}{Wagenländer et al.}

\begin{abstract}
Deep learning~(DL) jobs use multi-dimensional parallelism, \ie combining data, model, and pipeline parallelism, to use large GPU clusters efficiently. Long-running jobs may experience changes to their GPU allocation: (i)~resource elasticity during training adds or removes GPUs; (ii)~hardware maintenance may require redeployment on different GPUs; and (iii)~GPU failures force jobs to run with fewer devices. Current DL frameworks tie jobs to a set of GPUs and thus lack support for these scenarios. In particular, they cannot change the multi-dimensional parallelism of an already-running job in an efficient and model-independent way.

We describe \sys, a state management library for DL systems that enables jobs to change their parallelism dynamically after the GPU allocation is updated at runtime. \sys achieves this through a new abstraction, a \emph{parallelizable tensor collection}~(PTC), that externalizes the job state during training. After a GPU change, \sys uses the PTC to transform the job state: the PTC repartitions the dataset state under data parallelism and exposes it to GPU workers through a virtual file system; and the PTC obtains the model state as partitioned checkpoints and transforms them to reflect the new parallelization configuration.
For efficiency, \sys executes PTC transformations in parallel with minimum data movement between GPU workers.
Our experiments show that \sys enables DL jobs to support dynamic parallelization with low overhead.
\end{abstract}

\ccsdesc[500]{Computing methodologies~Distributed computing methodologies}
\ccsdesc[500]{Computing methodologies~Machine learning}

\keywords{distributed machine learning, resource changes}

\maketitle

\pagestyle{plain}


\section{Introduction}
\label{sec:intro}

Deep learning~(DL) has led to remarkable progress in many domains, including conversational AI~\cite{chatgpt}, natural language processing~\cite{DBLP:conf/naacl/DevlinCLT19, DBLP:conf/nips/BrownMRSKDNSSAA20, DBLP:conf/icassp/HayashiYIY0TTZT20}, computer vision~\cite{DBLP:conf/cvpr/HeZRS16,DBLP:conf/cvpr/SzegedyVISW16}, and recommender systems~\cite{280902, DBLP:conf/kdd/ZhouZSFZMYJLG18, DBLP:conf/ijcai/GuoTYLH17}. These advances, however, are due to the ever-increasing sizes of deep neural network~(DNN) models and training datasets: \eg OpenAI's GPT-3 language model has 175~billion parameters, requiring around 700\unit{GB} of memory for the model with 32~bit floating-point numbers~\cite{DBLP:conf/nips/BrownMRSKDNSSAA20}. Large DNN models, therefore, are trained in a distributed fashion with parallel hardware accelerators, such as GPUs~\cite{DBLP:conf/icml/RainaMN09}, NPUs~\cite{DBLP:journals/micro/EsmaeilzadehSCB13}, or TPUs~\cite{DBLP:conf/isca/JouppiYPPABBBBB17}.

Many organizations have invested in DL clusters with thousands of GPUs~\cite{philly-trace}, and DL jobs are deployed on GPUs using \emph{multi-dimensional} parallelism~\cite{hybrid_parallelisation, rajbhandari2020zero, 10.1145/3542929.3563487}. It combines data~\cite{DBLP:conf/nips/KrizhevskySH12}, pipeline~\cite{DBLP:conf/nips/HuangCBFCCLNLWC19, DBLP:conf/sosp/NarayananHPSDGG19}, and tensor parallelism~\cite{DBLP:conf/nips/DeanCMCDLMRSTYN12}, and these strategies are implemented either by \emph{model-specific libraries}, (Megatron-LM~\cite{megatron-lm}, DeepSpeed~\cite{deepspeed}), or \emph{deployment-time parallelizers} (Alpa~\cite{alpa}, Unity~\cite{unity}).

Due to their high cost~\cite{epoch2022trendsingpupriceperformance}, organizations must manage GPU clusters efficiently. Users submit training jobs with multi-dimensional parallelism to a \emph{DL job scheduler}~\cite{singularity, li2022aryl}, which allocates it to GPUs. An emerging requirement is that, due to the long-running nature of DL jobs, the original GPU allocation of a job may change over time~\cite{singularity} for several reasons: (i)~\textbf{elasticity}---to maintain high cluster utilization, DL jobs want to claim extra GPU resources when they become available~\cite{pytorch}; (ii)~\textbf{redeployment}---DL jobs may have to release specific GPUs and migrate to others to reduce fragmentation~\cite{reduce_fragmentation}, support hardware maintenance, or handle preemption by higher priority jobs~\cite{optimus}; and (iii)~\textbf{failure recovery}---DL jobs may lose GPUs at runtime due to failures and must continue training with fewer GPUs after recovering from checkpoints~\cite{DBLP:conf/sosp/WangJZZFNW23}.

We observe that current DL systems (PyTorch~\cite{pytorch}, TensorFlow~\cite{tensorflow}, MindSpore~\cite{mindspore}) do not allow DL job schedulers to change GPU resources at runtime. They lack a property that we term \emph{device-independence}: DL jobs are tightly coupled to GPUs at deployment time, preventing schedulers from changing the allocation. As we show in \S\ref{sec:background:challenges}, changing the GPU allocation of a job with multi-dimensional parallelism also means that its current parallelization strategy may no longer be optimal, thus requiring the replanning of its parallelization approach.

Both industry~\cite{singularity} and academia~\cite{kungfu} have recognized the need for changing DL job resources dynamically, resulting in three types of solutions: (a)~\textbf{model parallelizers}, \eg Megatron-LM~\cite{megatron-lm} and DeepSpeed~\cite{deepspeed}, can be extended with support to change GPU allocation during training. Such approaches, however, are limited to supported DNN architectures, such as specific transformer models~\cite{megatron-lm}; (b)~\textbf{elastic DL systems}, \eg Torch Elastic~\cite{pytorch_elastic}, Elastic Horovod~\cite{elastic_horovod}, and KungFu~\cite{kungfu} can adapt the number of model replicas on GPUs at runtime. By only adapting model replicas, such solutions are limited to data-parallel DL jobs only and do not support generic multi-dimensional parallelism; and (c)~\textbf{virtual devices}, used by \eg VirtualFlow~\cite{virtualflow}, EasyScale~\cite{easyscale}, and Singularity~\cite{singularity}, decouple DL jobs from physical devices: jobs assume a maximum number of virtual GPUs, which are then mapped to fewer physical GPUs at runtime. While this is transparent to job execution, it requires complex virtualization at the GPU driver level~\cite{singularity} and does not support changes with multi-dimensional parallelization~\cite{easyscale}.

In this paper, we explore a different point in the design space for supporting dynamic resource changes in DL clusters. Our idea is to create a \textbf{state management library} for DL systems that (i)~\emph{externalizes} the training state from a DL job (\ie the model and dataset partitions); and then (ii)~\emph{transforms} the state in response to dynamic GPU changes.

To design such a library, we answer several questions: (1)~what is a suitable abstraction for representing the DL job state, so that it can be transformed when adapting multi-dimensional parallelism after a GPU change? (2)~how can a state management library retrieve the job state from the DL system with little change to its implementation? (3)~how can the library deploy changes to the multi-dimensional parallelism of large DL jobs with low overhead?

We describe \textbf{\sys},\footnote{\url{https://github.com/kungfu-team/tenplex}} a state management library for DL systems that enables jobs with multi-dimensional parallelism to support dynamic changes to GPUs during training. \sys makes the following new technical contributions:

\mypar{(1)~Externalizing DL job state} \sys extracts the DL job state from the DL system and represents it using a tensor-based abstraction, which we call a \emph{parallelizable tensor collection}~(PTC). A PTC is a hierarchical partitioned collection of tensors that contains the (i)~\emph{dataset state} of the job, expressed as a set of training data partitions, and (ii)~the \emph{model state}, expressed as partitioned checkpoints of the DNN model parameters. The PTC partitioning depends on the multi-dimensional parallelization of the job, \ie how the job uses data, pipeline, and tensor parallelism.

\sys must expose the DL job state to a PTC and support efficient access by the DL system. \sys stores a PTC in a hierarchical virtual file system (implemented using Linux' FUSE interface~\cite{fuse}), which is maintained in memory for efficient access: (1)~for the dataset state, \sys loads the training data into the workers' host memory. To support data parallelism (DP), each data partition has a virtual directory. It contains the files with the training data samples that the worker must process; (2)~for the model state, \sys retrieves the partitioned model checkpoints created by the DL system, and the PTC stores them as a hierarchy of virtual files. The hierarchy mirrors the layered structure of the partitioned model tensors, simplifying state transformations when the multi-dimensional parallelization changes.

\mypar{(2)~Transforming DL job state} When the DL job scheduler alters the GPU allocation, \sys transforms the state maintained as a PTC to change the multi-dimensional parallelization configuration. After a GPU change, \sys requests a new parallelization configuration from a parallelizer (\eg Megatron-LM~\cite{megatron-lm} or Alpa~\cite{alpa}). It then applies \emph{state transformations} to the PTC that updates the partitioning of the tensors that represent the dataset and model states.

The state transformations ensure that the PTC remains consistent, \ie the convergence of the DL job is unaffected. For the dataset state, \sys repartitions the training data and makes the new data partitions available to workers while keeping the data access order of samples unaffected across iterations; for the model state, \sys repartitions the model layers and associated tensors and creates new partitioned model checkpoints. The partitioned checkpoints are then loaded by the new set of GPU devices.

\mypar{(3)~Optimizing DL job state changes} The reconfiguration of the DL job state must be done efficiently, \eg reducing data transfers to disseminate the new state to workers. \sys therefore parallelizes the PTC transformations across all workers, and it sends the minimum amount of data to establish  correctly partitioned state on all workers. Workers fetch sub-tensors from each other to avoid unnecessary data movement. \sys also overlaps the sending of dataset partitions with model training, which permits the DL system to resume training before the entire partitions are received.

\tinyskip

\noindent
We implement \sys as a Go library with 6,700~lines of code. It integrates with existing DL libraries, such as PyTorch~\cite{pytorch}, and systems, such as Megatron-LM~\cite{megatron-lm} and DeepSpeed~\cite{deepspeed}. Our evaluation shows that \sys can dynamically change a DL job's GPU resources with any parallelization configuration with good performance: it reduces training time by 24\% compared to approaches that only scale along the data parallelism dimension; resource reconfiguration takes 43\% less time than approaches that migrate all GPU state and 75\% less compared to maintaining state centrally.



\section{Resources Changes in DL Training}
\label{sec:background}

Next, we describe DL jobs with multi-dimensional parallelism (\S\ref{sec:DLtraining}). We then motivate resource changes during training~(\S\ref{sec:resourceChange}) and discuss associated challenges (\S\ref{sec:background:challenges}). We finish with a survey of current approaches for adapting resources during training and their limitations~(\S\ref{sec:SOTA}).

\subsection{Deep learning jobs with parallelism}
\label{sec:DLtraining}

Training DNN models, \eg large language models~(LLMs)~\cite{radford2019language}, is resource-intensive and must scale to clusters with many accelerators, such as GPUs~\cite{DBLP:conf/icml/RainaMN09}, NPUs~\cite{DBLP:journals/micro/EsmaeilzadehSCB13}, or TPUs~\cite{DBLP:conf/isca/JouppiYPPABBBBB17}. A single DL job may be executed on 1,000s of GPUs~\cite{DBLP:journals/corr/abs-2201-11990} by distributing it across workers, each with multiple GPUs.

\begin{figure}[tb]
  \centering
  \includegraphics[width=.55\columnwidth]{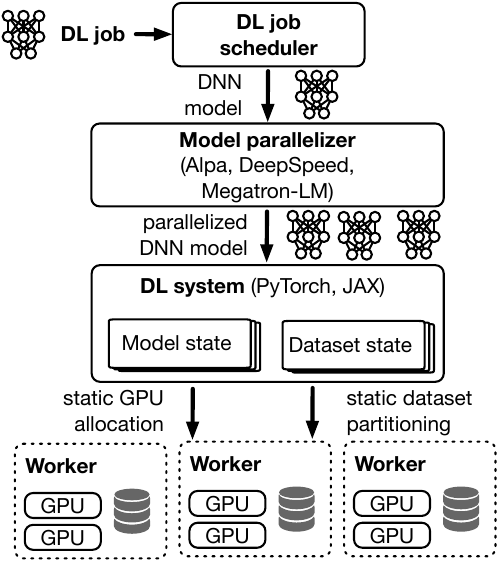}
  \caption{Training deep learning~(DL) jobs with multi-dimensional parallelism on a shared GPU cluster}
  \label{fig:design-current}
\end{figure}

\F\ref{fig:design-current} shows a typical deployment for DL jobs in a GPU cluster. A \emph{DL job scheduler} (\eg Pollux~\cite{pollux}, Gandiva~\cite{gandiva}) manages the GPUs and assigns jobs to them. When running a job, a \emph{model parallelizer} (\eg Alpa~\cite{alpa}, DeepSpeed~\cite{deepspeed}, Megatron-LM~\cite{megatron-lm}) decides on a parallelization configuration for the job by considering multiple dimensions: (1)~\emph{data parallelism}~\cite{DBLP:conf/nips/KrizhevskySH12} partitions training data across workers and replicates the DNN model. Workers compute model updates using their local data partitions and synchronize these updates after each training iteration; (2)~\emph{tensor parallelism}~\cite{DBLP:conf/nips/DeanCMCDLMRSTYN12} splits the model, \ie the operators and parameters in the computational graph~\cite{tensorflow, mindspore, pytorch}, and assigns partitions to workers; and (3)~\emph{pipeline parallelism}~\cite{DBLP:conf/nips/HuangCBFCCLNLWC19, DBLP:conf/sosp/NarayananHPSDGG19} partitions the model into stages~\cite{DBLP:conf/nips/HuangCBFCCLNLWC19}. Training data batches are then split into smaller micro-batches and pipelined across workers.

Recent advances~\cite{DBLP:conf/sc/NarayananSCLPKV21, megatron-lm, flexflow, DBLP:journals/corr/GoyalDGNWKTJH17} have shown that a combination of parallelism along these dimensions, \ie \emph{multi-dimensional parallelism}, improves the performance of large DNN model training. Different parallelization configurations have different properties in terms of their scalability, device utilization, and memory consumption: data parallelism alone cannot scale to large deployments due to its synchronization overheads and reliance on large batch sizes~\cite{limited_scaling_of_data_parallelism}; pipeline parallelism under-utilize devices due to pipeline bubbles~\cite{DBLP:conf/sosp/NarayananHPSDGG19}; and tensor parallelism incurs high communication overheads but must be used to fit large models that surpass GPU memory~\cite{model_parallelism}. By combining multiple strategies, model parallelizers~\cite{alpa, unity} navigate this trade-off space.

After generating a multi-dimensional parallelization plan, the DNN model training is performed on the GPU cluster by a DL system (\eg PyTorch~\cite{pytorch}, TensorFlow~\cite{tensorflow}, MindSpore~\cite{mindspore}). The deployed DL job consists of the \emph{dataset state} and \emph{model state}: the dataset state contains the partitions with the training data samples, and records the read positions across these partitions; the model state consists of the partitioned model and optimizer parameters. The partitioned state is assigned to the workers~(see~\F\ref{fig:design-current}).

\subsection{Need for dynamic resource changes}
\label{sec:resourceChange}

Since it may take hours, days, or weeks to run a single DL job, \eg training a large language model~(LLM)~\cite{checknrun, mlaas, DBLP:conf/ccgrid/NicolaeLWBDC20}, the DL job scheduler may change the GPU allocation of jobs at runtime. There are several reasons for this:

\mypar{(1)~Elasticity} DL job schedulers may \emph{elastically} increase and decrease the allocated GPUs for a job based on the available resources~\cite{antman, li2022aryl}. When a job completes, an elastic scheduler can re-allocate the freed-up GPUs to other jobs, \eg giving each job a fair share of GPUs~\cite{gandiva_fair}. Higher priority jobs submitted to the cluster may need to take GPU resources away from already-running jobs~\cite{DeepPool}.

In cloud environments~\cite{azure,gcp}, elastic schedulers can take advantage of differences in GPU pricing. When lower cost ``spot'' GPUs become available~\cite{spot_GPUs}, the scheduler may add them to existing jobs; when spot GPUs are preempted, jobs must continue execution with fewer GPUs. Elastic schedulers thus improve shared cluster utilization, reduce cost, and decrease completion times~\cite{evidence_why_elastic_scheduling_good}.

\mypar{(2)~Redeployment} DL job schedulers may reallocate jobs to a new set of GPUs for operational reasons. For example, before performing hardware maintenance or upgrades, a job may have to be shifted to a new set of GPUs at runtime.

The redeployment of jobs can also reduce fragmentation in the allocated GPUs. If a job uses GPUs spread across disjoint workers, communication must use lower-bandwidth networks (\eg Ethernet or InfiniBand) as opposed to higher-bandwidth inter-connects between GPUs (\eg NVLink). A scheduler may therefore change the GPU allocation of a job to de-fragment it onto fewer workers~\cite{reduce_fragmentation}.

\mypar{(3)~Failure recovery} Long-running jobs may lose GPU resources due to failures, caused by hardware faults, network outages, or software errors~\cite{checknrun, DBLP:conf/ccgrid/NicolaeLWBDC20}. After a fault, the job must continue execution after recovering from the last state checkpoint~\cite{checkfreq}. In some cases, the failed worker or GPUs can be replaced by new resources before resuming the job; in other cases, the job can resume with fewer GPUs, which affects its optimal parallelization configuration.

\subsection{Challenges when changing GPU resources }
\label{sec:background:challenges}

Changing GPUs for a job at runtime adds challenges:

\begin{figure}[t]
  \centering
  \begin{subfigure}[t]{0.49\columnwidth}
    \includegraphics[width=1.0\columnwidth]{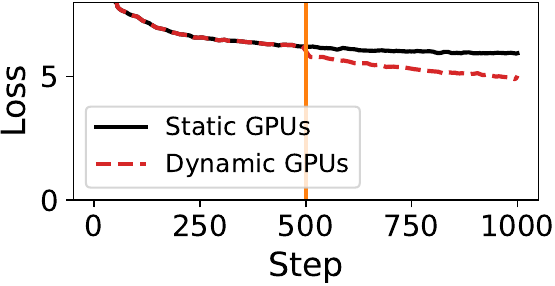}
    \captionsetup{aboveskip=1pt, belowskip=1pt}
    \caption{Inconsistent dataset access}
    \label{fig:change-mid-epoch}
  \end{subfigure}
  \begin{subfigure}[t]{0.49\columnwidth}
    \centering
    \includegraphics[width=\columnwidth]{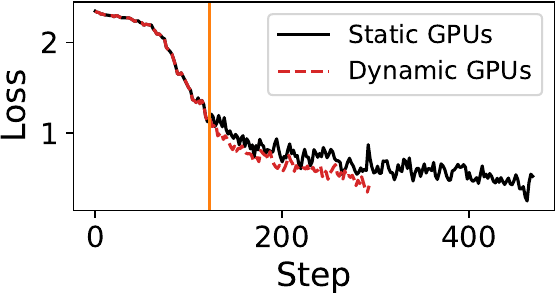}
    \captionsetup{aboveskip=1pt, belowskip=1pt}
    \caption{Inconsistent batch size}
    \label{fig:change-hyper}
  \end{subfigure}
  \caption{Impact of GPU change on training convergence \textnormal{(Changing GPUs from 2 to 4 with GPT-3 and MNIST)}}
  \label{fig:convergence}
\end{figure}

\begin{figure}[t]
  \centering
  \includegraphics[width=1.0\columnwidth]{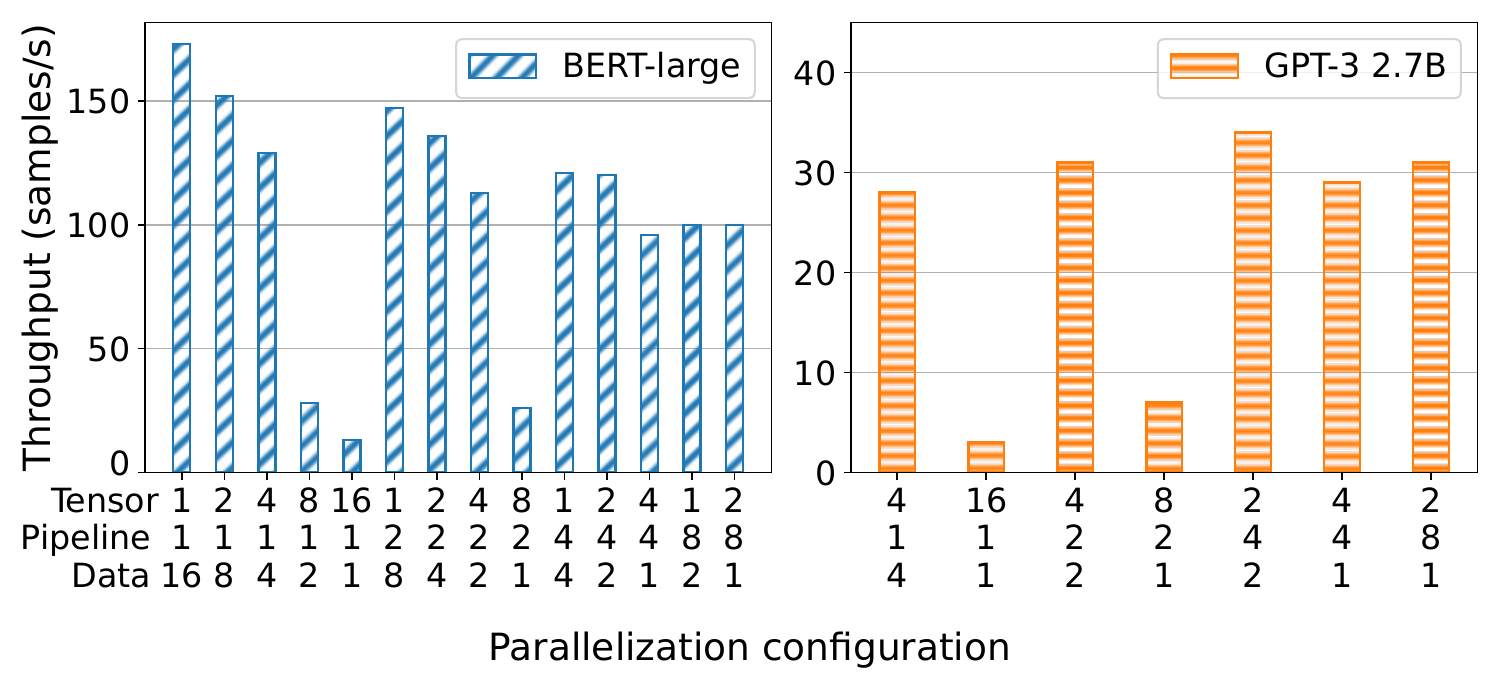}
  \caption{Performance impact of different parallelization configurations on 16 GPUs}\label{fig:mdp-config}
\end{figure}

\mypar{(1)~Impact on convergence} When changing the number of GPUs, the convergence of a job may be affected, and the final trained model may have \eg a different accuracy. In today's DL systems, job convergence depends on the specific set of GPUs used, as current jobs are not \emph{device-independent}. There are multiple reasons for this:

\mypari{Consistency of training dataset} A DL job must maintain dataset consistency during training, \ie it must process training data samples exactly once and in a consistent order in each training epoch. Dataset consistency must also hold when GPU changes under data parallelism, which affects the data sharding and requires re-partitioning. For example, when re-partitioning the dataset within an epoch, the order in which data samples are ingested from that point onwards must not change for convergence to be unaffected.

\F\ref{fig:change-mid-epoch} shows how model convergence, plotted as the loss value, is affected after adding a GPU (vertical orange line) under data parallelism. The solid black line shows regular model convergence with a static GPU allocation; the dashed red line shows convergence after the scale-out event when the dataset is processed inconsistently after re-partitioning: when resuming the training in the middle of the epoch, the first half of the training data is used twice, which overfits the model and reduces the loss value unreasonably.

\mypari{Consistency of hyper-parameters} Hyper-parameter choices, such as batch sizes, and learning rate~\cite{gradient-noise-scale}, depend on the GPU resources of a job. For example, the local batch size is fixed for each GPU and is typically chosen to keep devices fully utilized with data; the global batch size therefore changes with the number of GPUs.

In \F\ref{fig:change-hyper}, we show how the global batch size must be kept constant after adding a GPU (vertical orange line) under data parallelism. The solid black line shows model convergence (measured as loss) without the GPU change. The dashed red line shows the divergence when the GPU allocation changes but the device batch size remains constant.

\mypar{(2)~Impact on performance} The best parallelization configuration for a DL job, \ie one achieving the lowest time-to-accuracy, depends on the GPU resources used by the job.

\mypari{Parallelization configuration} The best multi-dimensional parallelization, in terms of data, tensor, and pipeline parallelism, depends on many factors, including the number and type of GPUs, the bandwidth and latency of the GPU inter-connect and the network between workers, and the size and structure of the DNN model architecture. Model parallelizers, \eg Alpa~\cite{alpa} and Unity~\cite{unity}, consider these factors based on profiled performance data and/or analytical cost models when choosing a parallelization configuration.

\begin{figure}[tb]
  \centering
  \includegraphics[width=.85\columnwidth]{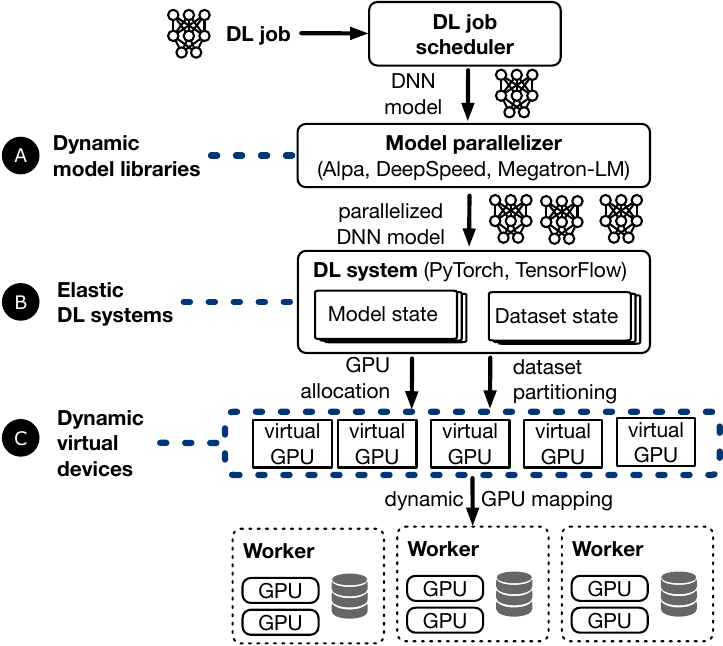}
  \caption{Approaches for dynamic resource changes in DL jobs}
  \label{fig:design-current_approaches}
\end{figure}

\begin{table*}[t]\sf\footnotesize\centering\setlength\extrarowheight{-5pt}
  \caption{Comparison of proposals for supporting dynamic GPU changes in DL jobs}
  \label{tab:comparison_approaches}
  \begin{tabular}{l@{~}p{2cm}lccc@{\hspace{2pt}}c@{\hspace{2pt}}cc@{\hspace{2pt}}c@{\hspace{2pt}}cc}
    \toprule
    & \multirow{3}{*}{\textbf{Approach}} & \multirow{3}{*}{\textbf{Systems}} & \multicolumn{2}{c}{\textbf{Consistency}} & \multicolumn{6}{c}{\textbf{Parallelism}} & \textbf{Reconfiguration} \\
    &&& \multirow{2}{*}{Dataset} & \multirow{2}{*}{Hyper-params} & \multicolumn{3}{c}{Static} & \multicolumn{3}{c}{Dynamic} & \textbf{overhead} \\

    &&&&& DP & PP & TP & DP & PP & TP & \\

    \midrule

    \multirow{4}{*}{\scell[l]{\myc{A}}} & \multirow{4}{*}{\scell[l]{Model libraries}} & Alpa~\cite{alpa} & - & - & \tickYes & \tickYes & \tickYes & - & - & - & - \\
    & & Megatron-LM~\cite{megatron-lm} & - & - & \tickYes & \tickYes & \tickYes & \tickYes & \tickNo & \tickNo & full state \\
    & & Deepspeed~\cite{deepspeed} & \tickYes & \tickYes & \tickYes & \tickYes &  \tickNo & \tickYes & \tickNo & \tickNo & full state \\

    \midrule

    \multirow{5}{*}{\scell[l]{\myc{B}}} & \multirow{5}{*}{\scell[l]{Elastic\\ DL systems}} & Elastic Horovod~\cite{elastic_horovod} & \tickNo & \tickNo & \tickYes & - & - & \tickYes & - & - & full state \\
    & & Torch Distributed~\cite{pytorch_elastic} & \tickYes & \tickNo & \tickYes & \tickYes & (\tickYes) & \tickYes & (\tickYes) & (\tickYes) & full state\\
    & & Varuna~\cite{varuna} & \tickYes & \tickYes & \tickYes & \tickYes & - & \tickYes & \tickYes & - & full state \\
    & & KungFu~\cite{kungfu} & \tickYes & \tickYes & \tickYes & - & - & \tickYes & - & - & full state \\

    \midrule

    \multirow{4}{*}{\scell[l]{\myc{C}}} & \multirow{4}{*}{\scell[l]{Virtual devices}} & VirtualFlow~\cite{virtualflow} & \tickYes & \tickYes & \tickYes & - & - & \tickYes & - & - & full state \\
    & & EasyScale~\cite{easyscale} & \tickYes & \tickYes & \tickYes & - & - & \tickYes & - & - & full state \\
    & & Singularity~\cite{singularity} & \tickYes & \tickYes & \tickYes & \tickYes & \tickYes & \tickYes & \tickNo & \tickNo & GPU state \\

    \midrule

    \multicolumn{2}{l}{State management} & Tenplex & \tickYes & \tickYes & \tickYes & \tickYes & \tickYes & \tickYes & \tickYes & \tickYes & minimal state \\
    \bottomrule
    \multicolumn{12}{l}{\rule{0pt}{10pt}
    \textnormal{\tickYes{} indicates support for the feature; (\tickYes) indicates support after a job-specific implementation by the user;}} \\
    \multicolumn{12}{l}{
    \textnormal{\tickNo{} indicates support but without dynamic scaling; and \textbf{-} indicates no support.}}
\end{tabular}
\end{table*}

When the GPU resources of a DL job change at runtime, a parallelization configuration that was optimal at deployment time may no longer be optimal with the new GPUs. We demonstrate this empirically in \F\ref{fig:mdp-config}, which shows the training throughput (in samples/second) when training BERT~\cite{DBLP:conf/naacl/DevlinCLT19} and GPT-3~\cite{DBLP:conf/nips/BrownMRSKDNSSAA20} models using Megatron-LM~\cite{megatron-lm} on 16~GPUs under a range of parallelization configurations (see \S\ref{sec:eval:setup}). Each parallelization configuration varies the degree of tensor, pipeline, and data parallelism, altering the GPU allocation.

As the results show, the training throughput differs by over 10$\times$ between the best and the worst configuration, despite the fact that each configuration uses the same number of GPUs~(16). The configuration~$(T,P,D) = (2,4,2)$ performs well because it uses communication-intensive tensor parallelism only within workers; the configuration~$(16,1,1)$ performs the worst because tensor parallelism must use slower inter-worker links.

\mypari{Reconfiguration cost} After changing parallelization, a job's partitioned dataset and model state are no longer correct. The state must be re-partitioned and the new partitions must be sent to workers, which may involve large data movement. For example, prior work~\cite{singularity} reports that reducing the GPU allocation for a DL job from 16 to 8 GPUs may take 122\unit{secs}.

\subsection{Current approaches}
\label{sec:SOTA}

A number of approaches have been proposed to allow DL job schedulers to change GPU resources dynamically. We give an overview, and then discuss specific proposals, assessing them against the challenges from \S\ref{sec:background:challenges}.

\F\ref{fig:design-current_approaches} shows the main approaches: \myc{A}~\emph{dynamic model libraries} adapt to changes in GPUs by producing a new parallelization configuration at runtime; \myc{B}~\emph{elastic DL systems} include support to scale GPU resources out and in at runtime; and \myc{C}~\emph{virtual devices} decouple physical from logical GPUs, allowing the mapping to change at runtime.

\T\ref{tab:comparison_approaches} compares systems that implement these approaches.

\mypar{\myc{A}~Model libraries} The Alpa model parallelizer~\cite{alpa} provides a parallelization configuration at job deployment time and does not support dynamic changes. Megatron-LM~\cite{megatron-lm} and DeepSpeed~\cite{deepspeed} support dynamic resource changes under data parallelism only by dividing batches into mini-batches~\cite{DBLP:journals/corr/abs-2201-11990, DBLP:conf/icml/RajbhandariLYZA22}. By changing the allocation of mini-batches, DeepSpeed ensures consistency after resource changes. Since the full training state is moved to and from remote storage, there is a high reconfiguration overhead.

In summary, model libraries do not handle dynamic multi-dimensional parallelism, and they lack integration with DL systems, requiring manual state re-partitioning.

\mypar{\myc{B}~Elastic DL systems} Elastic Horovod~\cite{horovod} exposes the model state through a user-defined state object. It allows users to synchronize state across workers when changing data parallelism, but state re-distribution must be implemented manually. In particular, the dataset state can become inconsistent if scaling does not occur at epoch boundaries. Torch Distributed Elastic/Checkpoint~\cite{pytorchdistributed} provides a model broadcast API to save/resume model checkpoints and allows users to implement re-partitioning operations. Users must ensure the consistency of hyper-parameters and perform the required data movement between workers. Varuna~\cite{varuna} lets users define cut-points at which the model pipeline is partitioned at runtime when resources change. KungFu~\cite{kungfu} uses a broadcast operation to distribute the training state and with it the model replicas.

Overall, elasticity support either does not account for full multi-dimensional parallelism or requires users to implement state re-partitioning and distribution manually.

\myparr{\myc{C}~Virtual devices} make DL jobs device-independent by virtualizing resources and allowing the mapping between virtual/physical resources to change at runtime. The set of virtual resources exposed to a job represents the maximum resources available at runtime. VirtualFlow~\cite{virtualflow} uses an all-gather operation to send the current training state to the new workers. To ensure dataset consistency, it follows exactly once semantics for data loading. EasyScale~\cite{easyscale} uses a thread abstraction and performs process snapshotting to capture state. Singularity~\cite{singularity} obtains the full GPU state through virtualization at the CUDA driver level.

As a consequence, virtual device approaches are effective at supporting dynamic data parallelism, which does not require re-partitioning, but they cannot support runtime changes with multi-dimensional parallelism.



\section{\sys Design}
\label{sec:design}

\sys{}'s goal is to support dynamic GPU changes of DL jobs while (i)~ensuring the consistency of the training result, (ii)~supporting arbitrary reconfiguration of jobs with multi-dimensional parallelism, and (iii)~maintaining a low reconfiguration overhead (see~\T\ref{tab:comparison_approaches}).

\begin{figure}[t]
  \centering
  \includegraphics[width=1.0\columnwidth]{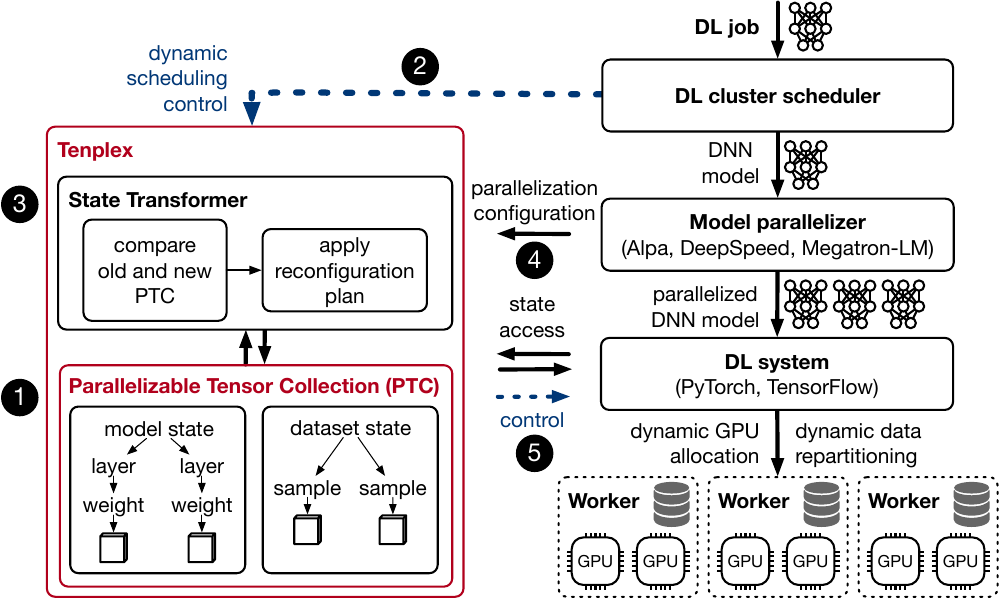}
  \caption{\sys design}\label{fig:design-overview}
\end{figure}

\sys's design is based on the observation that resource changes at runtime affect a DL job's state, but existing DL systems lack an abstraction to expose the state and transform it at runtime. \F\ref{fig:design-overview} shows the idea behind \sys{}'s design as a \emph{state management library} for DL systems. \sys \emph{externalises} the job state from the DL system~(see~\myc{1}) and manages it as a \emph{parallelizable tensor collection}~(PTC). A PTC provides a hierarchical tensor representation of the job's model and dataset state, and it enables \sys to modify the state across GPU devices after the parallelization configuration is updated, transparently to the DL system.

The updates to the GPU resources of a job are decided by a DL scheduler at runtime. The scheduler can increase or decrease a job's GPU allocation and notify \sys (\myc{2}). \sys then invokes its State Transformer~(\myc{3}), which first obtains a new parallelization configuration for the job from a model parallelizer, such as Alpa~\cite{alpa} or Megatron-LM~\cite{megatron-lm}~(\myc{4}). Based on the new configuration, the State Transformer calculates a \emph{reconfiguration plan}. The plan describes how the state represented by the PTC must change across the GPU devices to implement the new parallelization configuration. The reconfiguration plan is then executed by re-partitioning and re-distributing the data and model state~(\myc{5}).

We explain how the PTC abstraction allows \sys to manage the state of a DL job with multi-dimensional parallelism in \S\ref{sec:ptc}, and how \sys implements the state changes required by the reconfiguration plan efficiently in \S\ref{sec:impl}.


\section{Parallelizable Tensor Collection}
\label{sec:ptc}

Next, we describe the PTC abstraction~(\S\ref{sec:ptc:overview}) and how it is used by \sys to compute reconfiguration plans~(\S\ref{sec:ptc:reconfig}).

\subsection{PTC overview}
\label{sec:ptc:overview}

A \emph{parallelizable tensor collection}~(PTC) is \sys{}'s abstraction to represent the parallelized state of a DL job. Such an abstraction must satisfy several requirements: a PTC must match how the DL system represents parallelized state so that \sys can obtain and adapt the state correctly~(R1); it must represent the state of any multi-dimensional parallelization strategies to make \sys compatible with current parallelizers and their parallelization approaches~(R2); and it must facilitate the computation of an efficient reconfiguration plan for transforming the state from the current configuration to a new one, \eg with minimal data movement between GPU workers~(R3).

Since a PTC must capture the full execution state of a DL job, it includes (i)~the \emph{dataset state}, which consists of the training data and an iterator that records the processed data in the current epoch; and (ii)~the \emph{model state}, which consists of the DNN model parameters of all layers. A PTC expresses both types of state in a unified manner as a collection of \emph{tensors}, which allows \sys to manipulate the tensors when changing the parallelization configuration.

It is efficient for \sys to manage both types of state using the PTC~(R1): the dataset state is maintained directly in the PTC by \sys and exposed to the DL system through a suitable data access API; and the current model state is retrieved prior to reconfiguration from distributed model checkpoints created by the DL system~(see~\S\ref{sec:impl:tensor-store}).

\begin{figure}[t]
  \centering
  \includegraphics[width=1.0\columnwidth]{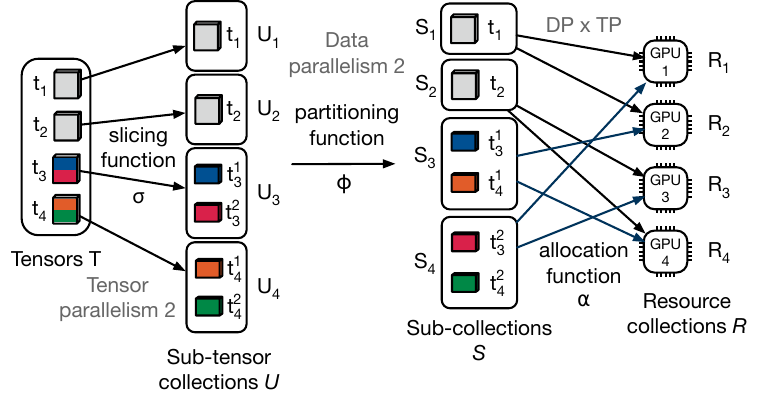}
  \caption{PTC example \textnormal{(Edges denote tensor mappings.)}}
  \label{fig:ptc-func}
\end{figure}

The PTC abstraction must be compatible with any multi-dimensional parallelism strategies~(R2). This allows \sys to support an arbitrary parallelizaton configuration that combines data, model, and pipeline parallelism~\cite{DBLP:conf/nips/HuangCBFCCLNLWC19, deepspeed, pathways}, as provided by model parallelization libraries (\eg Megatron-LM~\cite{megatron-lm}) or auto parallelizers (\eg Alpa~\cite{alpa}, Unity~\cite{unity}). \sys must then capture the impact of the parallelization strategy on the PTC state. Here, we observe that any multi-dimensional parallelization strategy can be expressed as a \emph{slicing} of state tensors, followed by a \emph{partitioning} of these tensors across GPU devices.

PTC exploits this observation to define parallelism with three mapping functions: (i)~a \emph{slicing} function encodes how tensors are split into sub-tensors, as dictated by tensor parallelism; (ii)~a \emph{partitioning} function then groups these sub-tensors into collections that can be assigned to devices, capturing data and pipeline parallelism; and (iii)~an \emph{allocation} function maps these sub-tensor collections to GPU devices for execution. Despite their simplicity, these three functions are sufficient to express any multi-dimensional parallelization strategies in a DL job.

We define the PTC as a tensor collection~$T$, slicing function~$\sigma$, partitioning function~$\phi$, and allocation function~$\alpha$:
\begin{equation}\label{eq:PTC}
	\ptc = (T, \sigma, \phi, \alpha)
\end{equation}
where $T = D \cup M$ are the tensors that make up the model~$M$ and dataset~$D$, $T = \{t_1, \dots, t_n\}$; $\sigma$ slices a tensor~$t$ into sub-tensors, $\sigma(t) = \{t^1, .., t^m\}, t \in T$. We denote all sets of sub-tensors as $U$, \ie $U=\{\sigma(t_1)\dots \sigma(t_{n})\}$; $\phi$ partitions $U$ into sub-collections~$S$, $\phi(U) = \{S_1, .., S_p\}$; $\alpha$ allocates sub-collections to GPUs of the resource pool~$R$,  $\alpha(S_i) = \{r_1, .., r_q\}$.

\F\ref{fig:ptc-func} shows an example of a PTC that describes the state of a job deployed on 4~GPUs with both model parallelism and data parallelism of degree 2. Here, the data samples~$\{t_1, t_2\}$ and model parameters~$\{t_3, t_4\}$ are tensors. With two-way tensor parallelism, each model tensor must be split into 2~sub-tensors: the slicing function~$\sigma$ slices each tensor in $T$ and creates a collection of sub-tensors, $U = \{U_{1}, \ldots, U_4\}$. With two-way data parallelism, each data tensor becomes its own sub-collection, and the model tensors are grouped by sub-tensor offset~$j$ of $t_i^j$: the partitioning function~$\phi$ takes $U$ and maps it to 4~sets. These sets are then assigned to the GPUs, $R_1$ to $R_4$, by the allocation function~$\alpha$, forming a cross-product of the data and model sub-collections.

\subsection{Reconfiguration plan}
\label{sec:ptc:reconfig}

The PTC abstraction allows \sys to decide how to reconfigure a DL job by computing a ``delta'' between the two PTCs: if there is a current \ptcc and a new \ptcn, it is possible to compute a minimal sequence of operations that must be executed to turn the state of \ptcc into that of \ptcn. We term such a sequence of operations a \emph{reconfiguration plan}.\looseness=-1

We observe that a reconfiguration plan can be expressed only in terms of \emph{split}, \emph{re-partition} and \emph{merge} operations. These operations update the data and model tensors in the PTC: the reconfiguration plan takes the sliced/partitioned tensors that exist on the GPU workers, as described by \ptcc, and transforms them, so they become the state described by \ptcn. This is done efficiently by only exchanging a minimal set of sub-tensors between GPU workers.

To generate such a reconfiguration plan, \sys considers the differences between the current PTC functions, ($\sigma$, $\phi$,  $\alpha$), and the new ones, ($\sigma'$, $\phi'$, $\alpha'$). It then generates a sequence of operations: (i)~if the sub-tensors~$U$ of \ptcc and $U'$ of \ptcn are different, a \emph{split} operation slices the sub-tensors according to the current slicing function~$\sigma$ and the new~$\sigma'$; (ii)~a \emph{re-partition} operations move the split tensors from a previous GPU~$R$ to~$R'$; and (iii)~if sub-tensors were previously split but are now on the same GPU, a \emph{merge} operation combines them again to reflect $\sigma'$. Performing the \emph{re-partition} operation between the \emph{split} and \emph{merge} operations minimizes data movement because only necessary tensors are moved.

\begin{figure}[t]
  \centering
  \includegraphics[width=1.0\columnwidth]{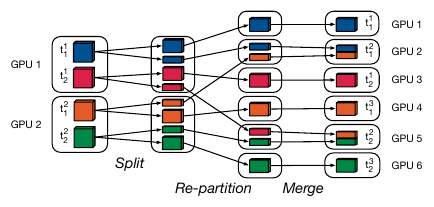}
  \caption{Reconfiguration plan \textnormal{(Edges denote tensor mappings.)}}\label{fig:transformation-slice-merge}
\end{figure}

\F\ref{fig:transformation-slice-merge} shows an example of a reconfiguration plan, which contains tensor parallelism~(TP) with 2~GPUs, to \ptcn, which combines tensor~(TP) and pipeline parallelism~(PP) with 6~GPUs. The tensors must change from a slicing into two sub-tensors, $\sigma(t_i) = \{t_i^1, t_i^2\}$, to a slicing, $\sigma'(t_i) = \{t_i^1, t_i^2, t_i^3\}$, into 3~sub-tensors. Therefore, the \emph{split} operation divides up each sub-tensor into two parts, which the \emph{re-partition} operation moves to new GPUs, as defined by the new partition function $\phi'$, forming the sub-tensor~$t_i^j$. In the final step, the \emph{merge} operation takes the split tensors and merges them into required sub-tensors. With PP of degree~2, there is only 1~tensor per stage, and with TP of degree~3, there is only one sub-tensor per GPU.


\SetAlFnt{\sffamily}
\SetArgSty{text}
\begin{algorithm}[t]
\footnotesize
\DontPrintSemicolon
\SetKwProg{Fn}{func}{}{}
\caption{Reconfiguration plan generation}
\label{alg:transformation}
    \KwData{$\ptc = (T, \sigma, \phi, \alpha)$, $\ptc' = (T, \sigma', \phi', \alpha')$ \\
            Resources $R, R'$\\
    }
    \KwResult{Reconfiguration plan $\mathcal{P}$}

	$U \gets \{\sigma(t) | \, t \in T\}$ \tcp*[f]{\textit{get sub-tensor collections}}\\
    \ForEach(\tcp*[f]{\textit{start SPLIT}}){$r \in R$}{
        $V \gets \{v | \, v \in U, \alpha(\phi(U)) = r \}$  \tcp*[f]{\textit{get sub-tensors of $r$}}\\ \label{alg:reconf:filter-sub-tensors}
        \ForEach{$v \in V$}{
            $\mathcal{P} \gets \mathcal{P} \mathbin\Vert$ split($v$, $\sigma$, $\sigma'$) \label{alg:reconf:split}
        }
    }
    $S' \gets \phi'(\{\sigma'(t) | \, t \in T\})$ \tcp*[f]{\textit{get sub-collections}}\\
    \ForEach(\tcp*[f]{\textit{start RE-PARTITION}}){$r' \in R'$}{
        $S'_r \gets \{S_i' | \, S_i' \in S', \alpha(S_i') = r \}$ \tcp*[f]{\textit{get sub-tensors of $r'$}}\\ \label{alg:reconf:resource-sub-tensors}
        \ForEach{$s' \in S'_r$}{
            $t \gets$ get\_base\_tensor($\sigma'$, $\phi'$, $s'$)\\ \label{alg:reconf:deduce-base-tensor}
            $W \gets$ get\_split\_tensors($t$, $\sigma$, $\sigma'$)\\ \label{alg:reconf:deduce-split-tensors}
            \ForEach{$w \in W$}{
                $r_w \gets$ get\_resource($\phi$, $\alpha$, $w$)\\
                $\mathcal{P} \gets \mathcal{P} \mathbin\Vert$ move($w$, $r_w$, $r'$) \tcp*[f]{add MOVE}\\ \label{alg:reconf:move}
            }
            $\mathcal{P} \gets \mathcal{P} \mathbin\Vert$ merge($W$) \tcp*[f]{add MERGE}\\ \label{alg:reconf:merge}
        }
    }
\end{algorithm}

\A\ref{alg:transformation} formalizes the computation of reconfiguration plan~$\mathcal{P}$ from \ptcc and \ptcn. First, \sys performs the \emph{split}: it starts by generating the current sub-tensor set~$U$ based on the slicing function~$\sigma$. For each resource~$r$, it filters the sub-tensors by the sub-tensors~$v$ that are on resource~$r$ (line~\ref{alg:reconf:filter-sub-tensors}). For each sub-tensor~$v$, it then adds a \emph{split} operation to the reconfiguration plan~$\mathcal{P}$ (line~\ref{alg:reconf:split}). Where to split is decided based on the current~$\sigma$ and new slicing~$\sigma'$. Second, \sys considers \emph{re-partition}: it starts with all sub-collections~$S'$ and filters them by resource~$r'$ (lines~\ref{alg:reconf:resource-sub-tensors}). For each sub-tensor, it gets the base tensor~$t \in T$ for the sub-tensor~$s'$ and how it is divided after the \emph{split} operation (lines~\ref{alg:reconf:deduce-base-tensor}--\ref{alg:reconf:deduce-split-tensors}). For each split tensor, it obtains the resource~$r_w$ for split tensor~$w$. It then appends a \emph{move} sub-operation to the reconfiguration plan~$\mathcal{P}$ to move $w$ from $r_w$ to~$r'$~(line~\ref{alg:reconf:move}). Finally, \sys adds a \emph{merge} operation to~$\mathcal{P}$, which merges the split tensors in~$W$~(line~\ref{alg:reconf:merge}).

\subsection{Expanding to new parallelism strategies}
\label{sec:ptc:new-para}

In addition to data, tensor, and pipeline parallelism, the PTC abstraction is designed to accommodate new parallelism strategies. The generality of the $\ptc$ functions means that they can accommodate the parallelization and distribution of other strategies, such as expert parallelism~\cite{DBLP:conf/icml/RajbhandariLYZA22} in a mixture of experts~(MoE) settings~\cite{moe}, and sequence parallelism\cite{sequence-parallelism}:

Expert parallelism~(EP) can be expressed by using the partitioning function~$\phi$ to group expert-specific tensors together. Analogous to pipeline parallelism, groups of model tensors form a partition and then are allocated by the allocation function~$\alpha$ to GPUs, \ie replacing pipeline stage tensors with model expert tensors. Since EP does not split tensors, the slicing function~$\sigma$ is the identity function.

Sequence parallelism~(SP) uses the slicing function~$\sigma$ to slice data sample tensors along the sequence dimension. This makes it similar to tensor parallelism, but, instead of splitting the model tensors, it splits the data sample tensors.


\section{\sys Architecture}
\label{sec:impl}

In this section, we describe \sys{}'s architecture and how it implements the PTC abstraction to reconfigure DL jobs efficiently after resource changes.


As shown in \F\ref{fig:architecture}, \sys executes on each worker and has two main components: a distributed \emph{State Transformer} and an in-memory \emph{Tensor Store}. The State Transformer inputs the model and dataset partitions from a previous $\ptc$ and creates updated partitions to comply with a new $\ptc'$ after a resource change; the Tensor Store maintains the model and dataset state partitions represented by the PTC in a hierarchical virtual in-memory file system. It offers APIs to interface with the DL system's support for model checkpointing and to allow the DL system to ingest training data.

\subsection{State Transformer}
\label{sec:impl:state-transformer}

When the resources of a DL job change, \sys must create new state partitions based on the updated parallelization plan so that the DL system can resume executing the job. Since this state transformation can be parallelized, \sys maintains an instance of the \emph{State Transformer} for each resource~$r$ on a worker. Each State Transformer instance then applies its part of the reconfiguration plan~(see~\S\ref{sec:ptc:reconfig}).

\begin{figure}[t]
  \centering
  \includegraphics[width=1.0\columnwidth]{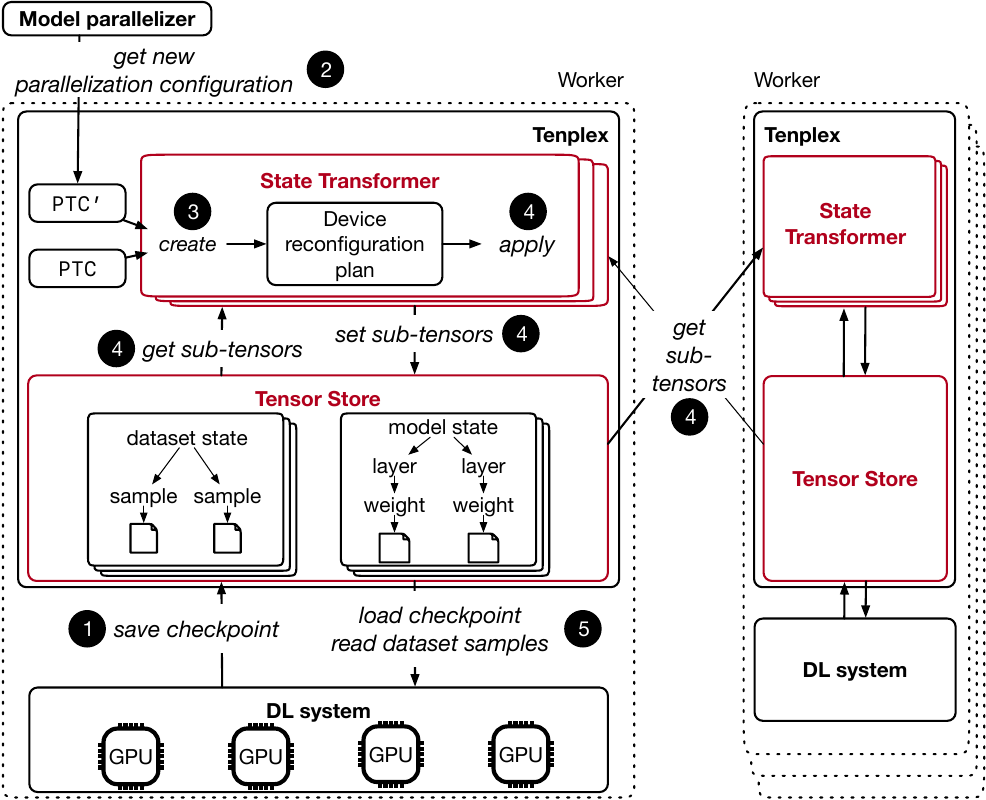}
  \caption{\sys architecture}
  \label{fig:architecture}
\end{figure}

\sys executes the following steps to modify the job after a resource change from the scheduler~(see~\F\ref{fig:architecture}): \myc{1}~\sys obtains the training state from the DL system by retrieving a model checkpoint partition per GPU. Each checkpoint is written to a model state partition in the Tensor Store; \myc{2}~it then requests a new parallelization configuration from the model parallelizer. The new configuration is expressed as $\ptc'$ and becomes the basis for the reconfiguration; \myc{3}~each State Transformer instance uses \A\ref{alg:transformation} to create the device's reconfiguration plan. It compares $\ptc$ and $\ptc'$ and infers the local transformation operations; \myc{4}~the State Transformer then applies the \emph{split}, \emph{re-partition}, and \emph{merge} operations to generate new state partitions~(see~\S\ref{sec:ptc:reconfig}). It retrieves the necessary sub-tensors from either the local or remote Tensor Stores and saves them in the local Tensor Store; and \myc{5}~it instructs the DL system to restore the job from the checkpoints based on the transformed model partitions in the local Tensor Store. After resuming the job, the DL system continues reading data samples from the local Tensor Store.

The State Transformer is designed to interact with different model parallelizers and thus provides a universal method for describing parallelization configurations. It processes these configurations as JSON objects with the following structure: the top level is a list of objects where each follows the structure of the model that a single GPU hosts, with the tensor shapes of the model parameters as leaves. From these objects, the \ptcc and \ptcn can be constructed.

\subsection{Tensor Store}
\label{sec:impl:tensor-store}

Each worker has an in-memory \emph{Tensor Store} that contains the model and dataset partitions for each local GPU. The Tensor Store maintains the model and dataset tensors in a hierarchical in-memory file system. The tree hierarchy follows the model structure, with model parameters and dataset samples as leaves.

\mypar{Model state} The Tensor Store exposes an API to the State Transformer to apply a reconfiguration plan, and to the DL system to load/store the model state. It supports a NumPy-like array interface~\cite{numpy} for requesting tensors via a REST API: a \texttt{query} request with a \texttt{path} attribute obtains a tensor; an \texttt{upload} request adds a new tensor at a given \texttt{path}.

A unique feature of the API is that it can be used to request \emph{sub-tensors} by defining a range for each dimension, similar to a Python slice. Requesting sub-tensors is important for performance, \eg when re-slicing under tensor parallelism, because it reduces data movement between workers. The State Transformer can request sub-tensors instead of complete tensors that would have to be split after transfer. To obtain a sub-tensor, a query includes a \texttt{range} attribute whose value specifies the dimension and range. For example, for a query that slices the second tensor dimension, the attribute is \texttt{range=[:,2:4]}, which returns the sub-tensor for $[2, 4)$.

The Tensor Store uses a simple API to move the model state in and out of the DL system. To obtain the current model state, \texttt{tenplex.save(model, path)} maps a Python dictionary with the model state to its hierarchical representation; \texttt{tenplex.load(path)} maps it back to a Python dictionary that can be consumed by the DL system.

The model state in the Tensor Store is represented as a hierarchical tree with node grouping parameters. For example, \texttt{``/2/embedding/weight''} is a weight parameter in the embedding layer of the model state partition~$2$. The leaves are sub-tensors, which are implemented as NumPy arrays to offer compatibility with most DL systems.

\mypar{Dataset state} The training dataset consists of binary files with data samples, which either reside on the local disk or remote storage. Data samples are tensors and \sys represents them as NumPy~\cite{numpy} (\code{npy} or \code{npz}) arrays.

\sys also maintains a \emph{dataset index} that maintains the locations of all data samples. Specifically, for each data sample, it holds the paths to the binary files and byte ranges within those files. Based on the dataset index, the State Transformer can repartition the dataset as necessary into partitions, each with its own indices. To read a data sample, the DL system invokes a data loader, which uses the partition-specific index to decide on the relevant binary file and byte range, reading the corresponding part of the file. The dataset index is stored in memory and uses 64-bit integer pairs for the byte range of each sample.

In contrast to the model state, the dataset is immutable and consumed sequentially. \sys leverages this to improve performance by overlapping training and dataset fetching. It streams the dataset into the Tensor Store on the workers while training iterations take place. Since the data sample order is known at the beginning of an epoch, \sys derives which samples to fetch first to unblock training.


In a typical cloud deployment, the training dataset is stored on remote storage, \eg S3~\cite{s3} or other blob stores~\cite{azure, gcp}. For training to resume, the data must be accessible by the workers, but the network bandwidth to remote storage is typically lower than the inter-worker bandwidth~\cite{DBLP:conf/sosp/WangJZZFNW23}. To reduce the impact of this, \sys tracks the location of data samples in the dataset index and distinguishes between remote and locally available data samples. It then prioritizes fetching samples from other workers and only uses remote storage if samples are otherwise unavailable.

\subsection{Fault tolerance}

When workers or GPU devices fail during job execution, \sys relies on the DL system to recover. After a failure, the job state is restored from the persisted checkpoints created by the DL system. Since a failure can be seen as a resource reduction, \sys{}'s reconfiguration support can be used to resume a job immediately, without waiting for new GPU resources. \sys thus resumes the job with fewer GPU devices but with an optimal new parallelization plan.

A deployment with \sys is subject to the usual trade-off between checkpointing frequency and overhead: with infrequent checkpointing, some job progress is lost after failure. \sys tries to avoid re-executing training steps due to stale checkpoints: for DL jobs with data parallelism, \sys exploits that the model state is replicated among workers. As long as at least one model replica remains after a failure, the state can be retrieved from that GPU.

To accommodate frequent failures in a cluster during training, \sys can replicate the model state in the Tensor Store across workers in a round-robin fashion, adding more state redundancy to the job. To obtain $n$~replicas, the state is replicated to the Tensor Stores of the next $n$~workers. If a worker fails and the state in the worker's Tensor Store is lost, the state can be recovered from another worker.

\subsection{Integration with existing training jobs}
\label{sec:architecture:integration}

To integrate with existing training jobs, \sys assumes that components support specific features. \sys makes the following assumptions about the DL software stack:

\myparr{Job schedulers} (\eg Kubernetes~\cite{kubernetes}, Pollux~\cite{pollux} , or Ray~\cite{DBLP:conf/osdi/MoritzNWTLLEYPJ18}, Sia~\cite{sia}) make resource management decisions according to a metric, such as shortest job first or priority. A job scheduler sends to \sys the DL model and information about GPU resources. If a resource change happens, the scheduler notifies \sys about the new resources. Based on this information, \sys orchestrates the reconfiguration and informs the scheduler after the reconfiguration is completed. If the number of resources is reduced, the scheduler can re-allocate the newly available resources.

\myparr{Model parallelizers} (\eg Alpa~\cite{alpa} and Megatron-LM~\cite{megatron-lm}) receive the DL model and the new resources from \sys and decide on the parallelization configuration. The parallelization configuration is expressed as a JSON object with the rank-specific model structure and tensor shapes. It describes how the data and model tensors are partitioned and mapped to GPUs (\S\ref{sec:impl:state-transformer}). \sys uses the parallelization configuration to construct the $\ptc$. 

\myparr{DL systems} (\eg PyTorch~\cite{pytorch} and JAX~\cite{jax}) must allow \sys to externalize the DL job state, \eg through APIs for extracting/loading model state to/from GPUs. The state must be expressed as per GPU Python dictionaries with the model state. For PyTorch, \texttt{model.state\_dict()} extracts the state and \texttt{model.load\_state\_dict()} loads the state.

\myparr{Training programs} must use \sys{}'s Tensor Store path when accessing the file system for data ingestion. In addition, users must replace calls for saving/loading checkpoints with \sys{}'s corresponding functions (\S\ref{sec:impl:tensor-store}). In the training loop, the training program must also invoke the \sys API to check if a reconfiguration is needed. If so, the training program must terminate, and \sys re-invokes the training program with the new resources after having transformed the job state.


\section{Evaluation}
\label{sec:eval}

We evaluate \sys with three use cases: supporting elastic scaling with \mdp~(\S\ref{sec:eval:elastic}), enabling job redeployment~(\S\ref{sec:eval:redeployment}), and handling failure recovery~(\S\ref{sec:eval:failure}). After that, we investigate \sys{}'s reconfiguration overhead~(\S\ref{sec:eval:diff-reconfiguration}), the impact of different parallelization strategies~(\S\ref{sec:eval:impact-model-size}), and its scalability in terms of cluster size~(\S\ref{sec:eval:impact-resource-count}). We finish by exploring \sys{}'s effect on model convergence when changing parallelism~(\S\ref{sec:eval:convergence}).

\subsection{Experimental setup}
\label{sec:eval:setup}

\begin{figure}[tb]
  \centering
  \includegraphics[width=1.0\columnwidth]{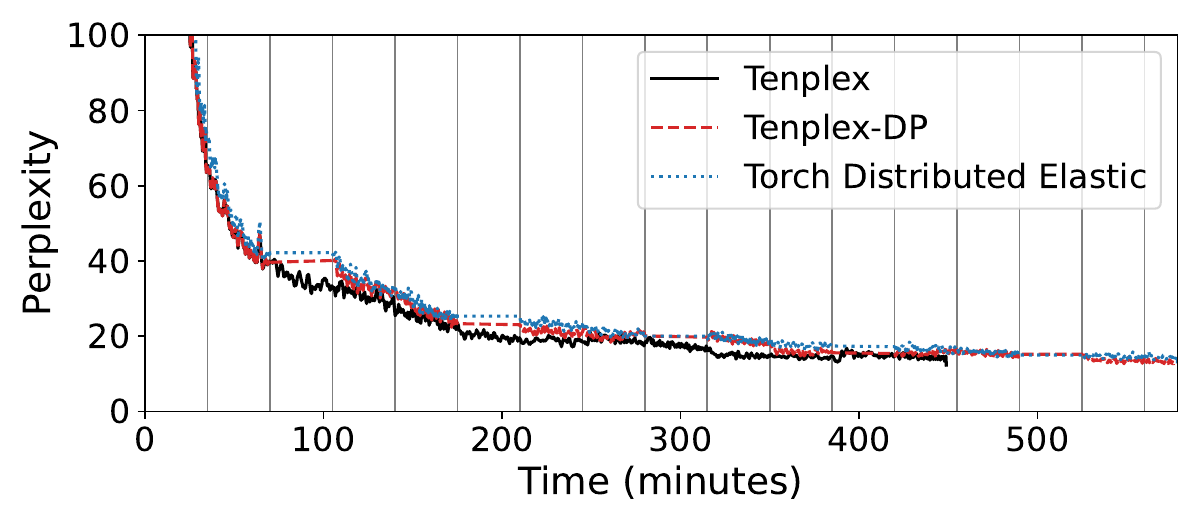}
  \caption{Elastic DL job convergence with \mdp under dynamic GPU changes}\label{fig:dynamic-resources}
\end{figure}

Our experiments have the following setup:

\mypar{Cluster} We conduct on-premise experiments with 16~GPUs (4~machines with 4~GPUs each). Each machine has an AMD EPYC 7402P CPU, 4 $\times$ NVIDIA~RTX A6000 GPUs, and PCIe 4.0. The machines are interconnected by 100-Gbps InfiniBand, and the GPUs are connected pairwise using 3rd generation NVLink. For experiments with a larger cluster, we also conduct 32-GPU cloud experiments on Azure~\cite{azure} with Standard\_NC24s\_v3 VMs, each with 4~NVIDIA V100 GPUs.

\mypar{Baselines} We compare \sys to multiple external baselines: (i)~\textsf{Torch} Distributed Elastic~v2.0 and (ii)~\textsf{Horovod}-Elastic~v0.28~\cite{elastic_horovod}, which are state-of-the-art elastic DL systems; and (ii)~\textsf{DeepSpeed}~v0.6~\cite{deepspeed} with \textsf{Megatron-LM}~v23.06, which represents the model library approach. All of these solutions can only support dynamic reconfiguration of DL jobs while changing the degree of data parallelism. We therefore also compare with (iv)~\textsf{\sys{}-DP}, which only reconfigures data parallelism, and (v)~\textsf{\sys{}-Central} that performs all state repartitioning at a central node.

\mypar{Models and datasets} We use these representative DNN models: (i)~\emph{BERT-large} with 340M parameters; (ii)~\emph{GPT-3} with 1.3B~(XL), 2.7B, and 6.7B parameters; and (iii)~\emph{ResNet-50} with 25M parameters.
For the training data, we use: (i)~\emph{OpenWebText}~\cite{openwebtext} with 2M samples with a sequence length of 1024; (ii)~\emph{Wiki\-pedia}~\cite{wikidump} with 6.8M samples and the same sequence length; and (iii)~\emph{ImageNet}~\cite{ImageNet} with 1M samples.

\subsection{Elastic \mdp}
\label{sec:eval:elastic}

\begin{figure*}[t]
  \minipage{0.32\textwidth}
  \includegraphics[width=.9\columnwidth]{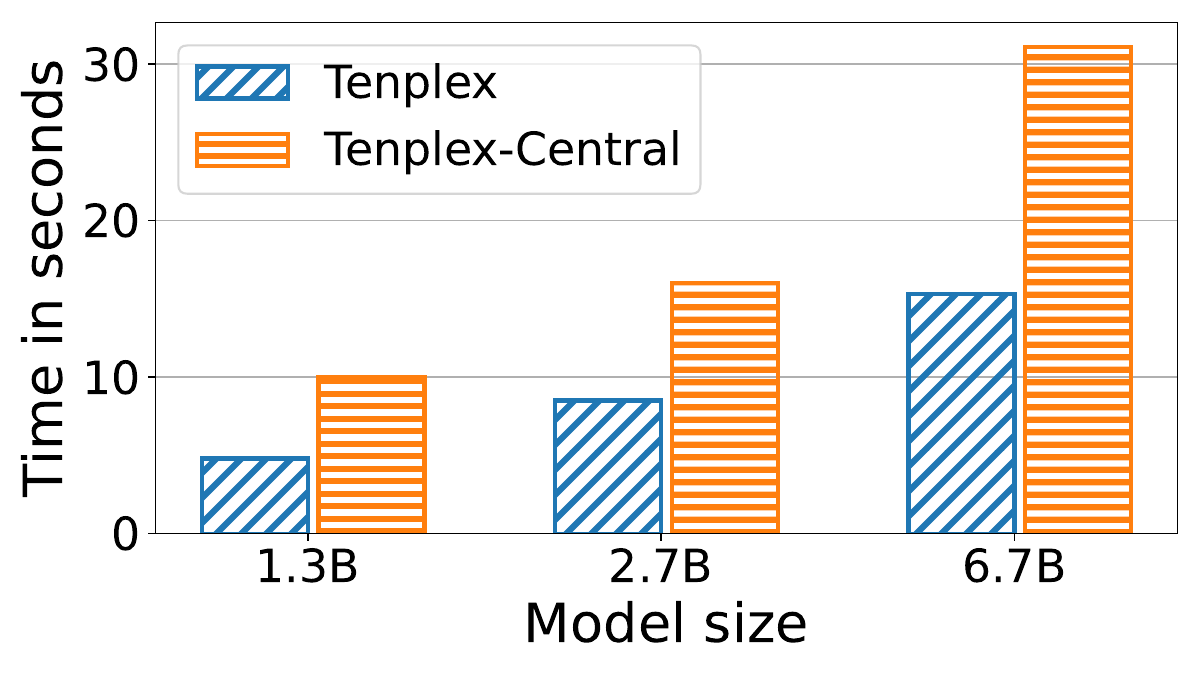}
  \caption{Redeployment time of DL job}
  \label{fig:redeployment}
  \endminipage
  \hfill
  \minipage{0.32\textwidth}
  \includegraphics[width=.9\columnwidth]{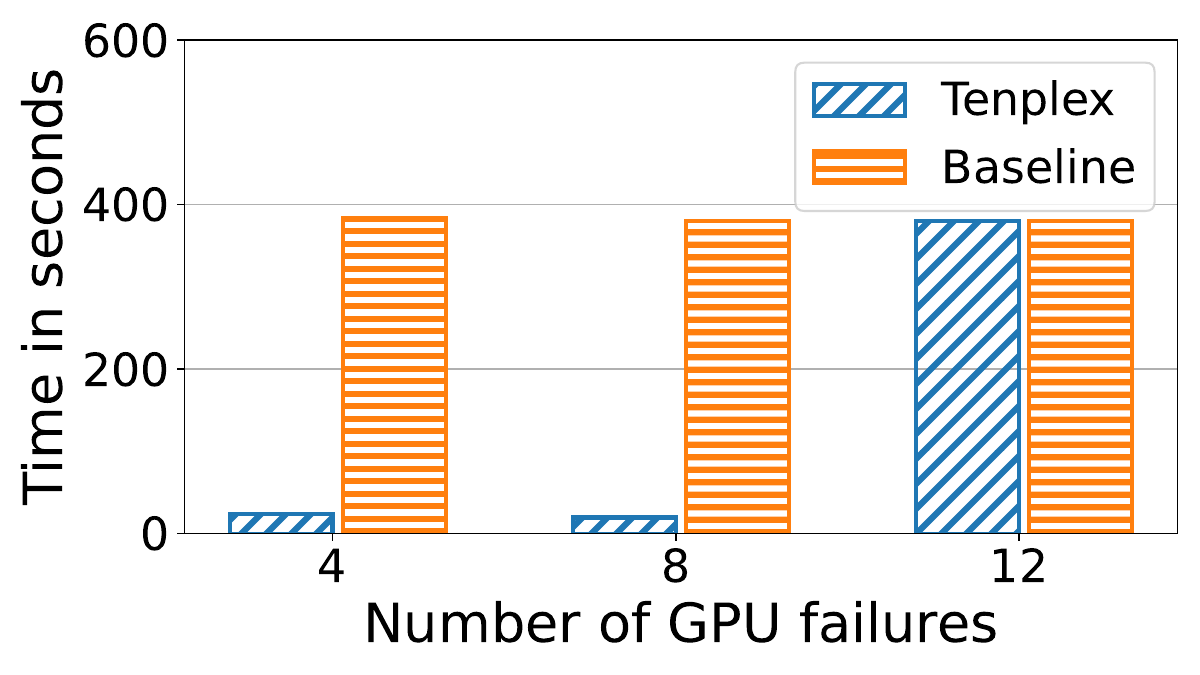}
  \caption{Failure recovery time \textnormal{(GPT-3 2.7\unit{B})}}
  \label{fig:failure-recovery}
  \endminipage
  \hfill
  \minipage{0.32\textwidth}
  \includegraphics[width=.9\columnwidth]{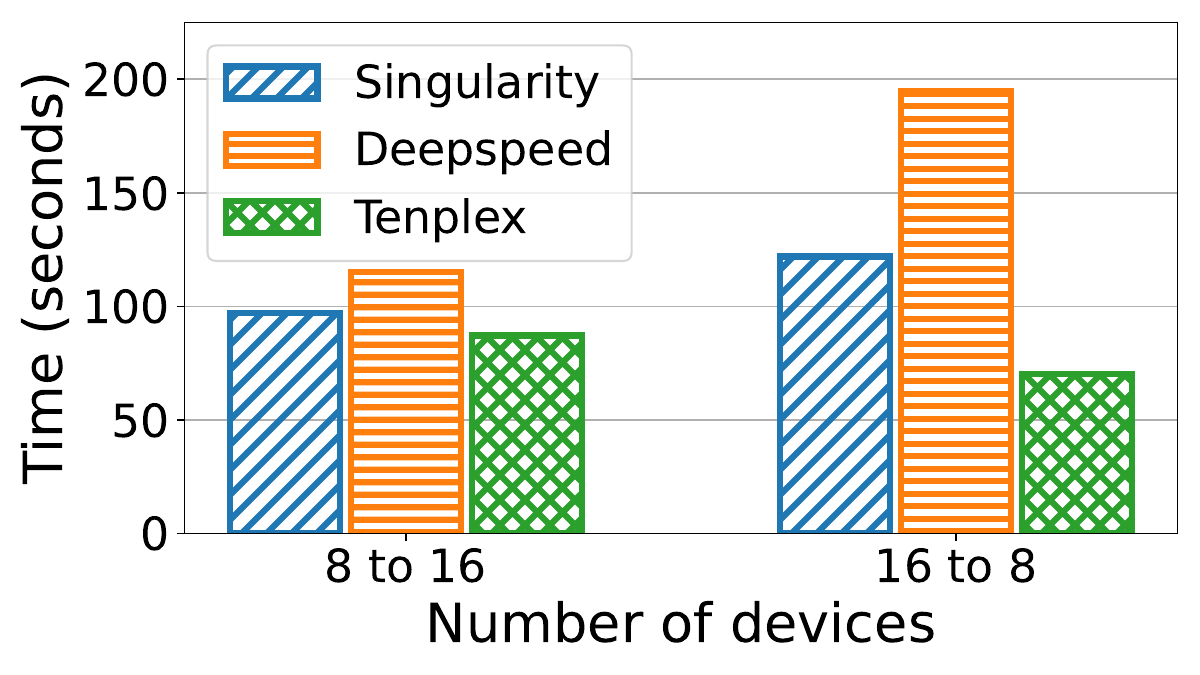}
  \caption{Reconfiguration time}
  \label{fig:comparison-scaling-latency}
  \endminipage
\end{figure*}

\begin{figure}[t]
  \centering
  \includegraphics[width=.7\columnwidth]{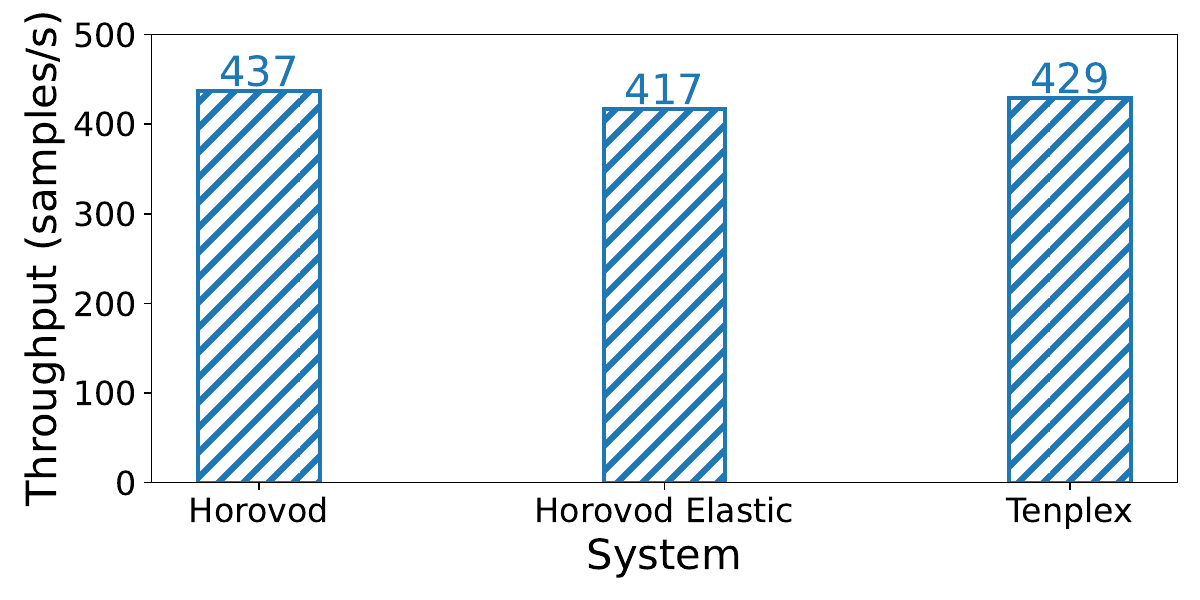}
  \caption{Training throughput against Horovod}
  \label{fig:throughput}
\end{figure}

First, we explore the benefits of supporting elasticity in DL jobs with \mdp, scaling across all parallelism dimensions when the GPU allocation changes.

In this experiment, we train DL jobs with the GPT-3 XL model on the on-premise 16-GPU cluster. The job runtime and elastic scaling events are derived from Microsoft's Philly trace~\cite{philly-trace}: over the runtime of 538\unit{mins}, we scale based on the average every 35\unit{mins}. During a scaling event, we change the number of GPUs for a job between 16, 8, and 4~GPUs.

We compare the training convergence of \sys to \textsf{\sys-DP} and \textsf{Torch} Distributed Elastic. \textsf{\sys-DP} is similar to \tde{} with Megatron-LM by only scaling dynamically along the data parallelism dimension.

In terms of the scaling decisions, \sys reconfigures the (tensor, pipeline, data) parallelism from $(T,P,D) = (2,4,2)$ to $(2,4,1)$ to $(2,2,1)$, which are the parallelization configurations that achieve the best performance. \textsf{\sys{}-DP} and \textsf{Torch} scale $(T,D,P) = (2,4,2)$ to $(2,4,1)$, and pause training with $4$~GPUs, because a configuration with pipeline parallelism of $4$ and tensor parallelism of $2$ cannot run on $4$~GPUs.

\F\ref{fig:dynamic-resources} shows the perplexity over time, and the scaling events are indicated as grey vertical lines, annotated by the GPU count. As we can see, \sys only takes 298~mins to reach the same step that \textsf{\sys-DP} reaches after 576~mins and \textsf{Torch} after 548~mins---a reduction by 46\%. Since \sys can support scaling along all dimensions in DL jobs, it exploits more optimal parallelization configuration using the GPU resources more effectively.

\subsection{Job redeployment}
\label{sec:eval:redeployment}

Next, we evaluate how long \sys takes to redeploy DL jobs with different model sizes onto a new set of GPU resources. As a baseline, we compare against \textsf{\sys{}-Central}, which follows the approach of PyTorch Elastic~\cite{pytorch_elastic} or DeepSpeed~\cite{deepspeed}: it holds all DL job state at a single central worker. In this experiment, we therefore specifically explore the benefit of \sys{}'s distributed state management.

We redeploy a DL job with \mdp from one set of 8~GPUs to another 8~GPUs. We measure the redeployment time on the on-premise clyster with the GPT-3 model with sizes of 1.3\unit{B}, 2.7\unit{B}, and 6.7\unit{B}. The parallelization configuration is $(T,D,P) = (4,2,1)$ and remains the same because the number of GPUs remains unchanged.

\F\ref{fig:redeployment} shows the redeployment time under different model sizes. In all cases, \sys achieves a lower redeployment time than \textsf{\sys{}-Central}: the time for \textsf{\sys{}-Central} is 2.1$\times$ for the 1.3\unit{B} model, 1.9$\times$ for 2.7\unit{B} model, and 2$\times$ for 6.7\unit{B} model, respectively, higher compared to \sys. With its distributed state management between State Transformer instances on different workers, \sys{} can migrate state directly between workers. This prevents the network bandwidth of any single worker from becoming a bottleneck, which would increase redeployment time.

\subsection{Failure recovery}
\label{sec:eval:failure}

We explore how \sys manages to recover from failures, even in scenarios that require dynamic reconfiguration due to a change in the number of GPUs. We assume a fail-stop failure model in which only the GPUs fail. We emulate faults of 4, 8, and 12~GPUs and measure the failure recovery and reconfiguration time. We use the GPT-3 2.7\unit{B} model with the Wikipedia dataset on the on-premise cluster. We compare \sys to a system that always recovers from the last checkpoint (denoted as Baseline), which results in an average loss of 50~training steps. The parallelization configuration is $(T,P,D) = (4,2,2)$, \ie there are two model replicas.

\F\ref{fig:failure-recovery} shows the recovery time in seconds with different numbers of failed GPUs. \sys recovers faster than the baseline if there exists at least one model replica, \ie for failures with 4 and 8~GPUs. Here, \sys does not need to rerun the lost training steps, because it does not rely on the stale checkpointed state for recovery. With $8$~GPUs, \sys takes only $5$\% of the recovery time of the baseline and exhibits the same cost as for 12~GPUs.

When there is no redundant model replica available, \sys uses the last checkpoint and only achieves a slight performance benefit over the baseline. This is due to using local storage instead of remote storage when recovering from the checkpointed state. We conclude that \sys reduces failure recovery times when model replicas are available due to the parallelization configuration.

\begin{figure*}[t]
  \centering
  \begin{subfigure}[t]{0.33\textwidth}
    \includegraphics[width=.9\columnwidth]{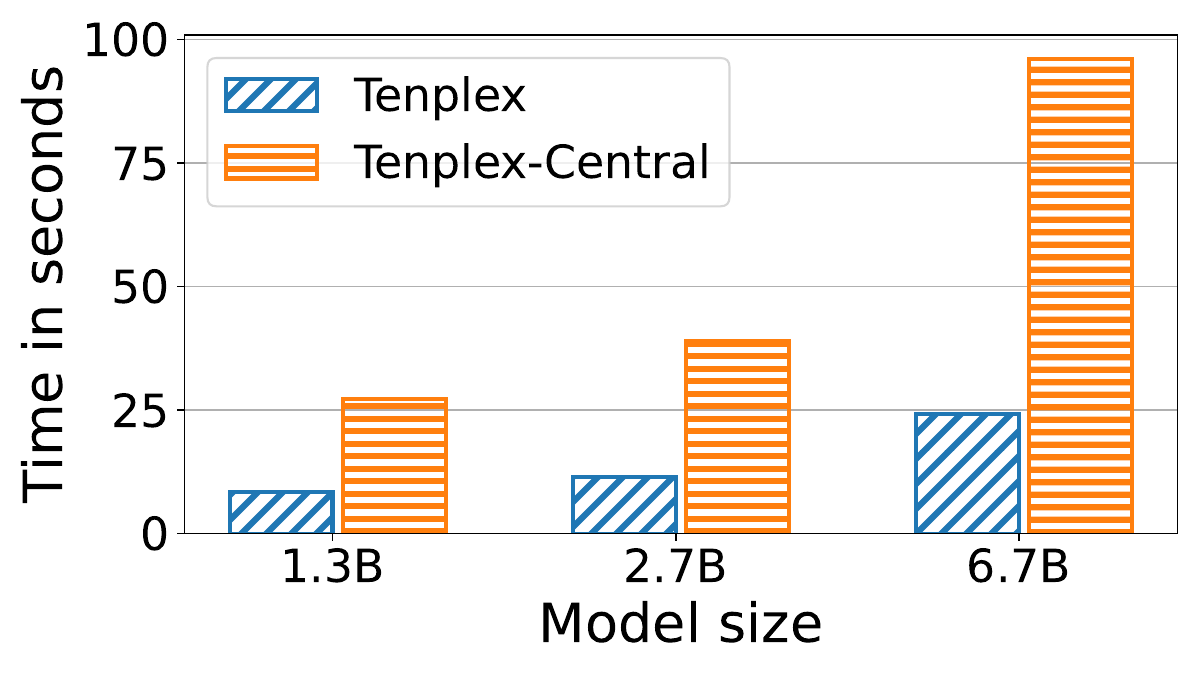}
    \captionsetup{width=.9\linewidth, aboveskip=1pt, belowskip=1pt}
    \caption{Data parallelism}
    \label{fig:state-transformer-time-size-dp}
  \end{subfigure}
  \hfill
  \begin{subfigure}[t]{0.33\textwidth}
    \includegraphics[width=.9\columnwidth]{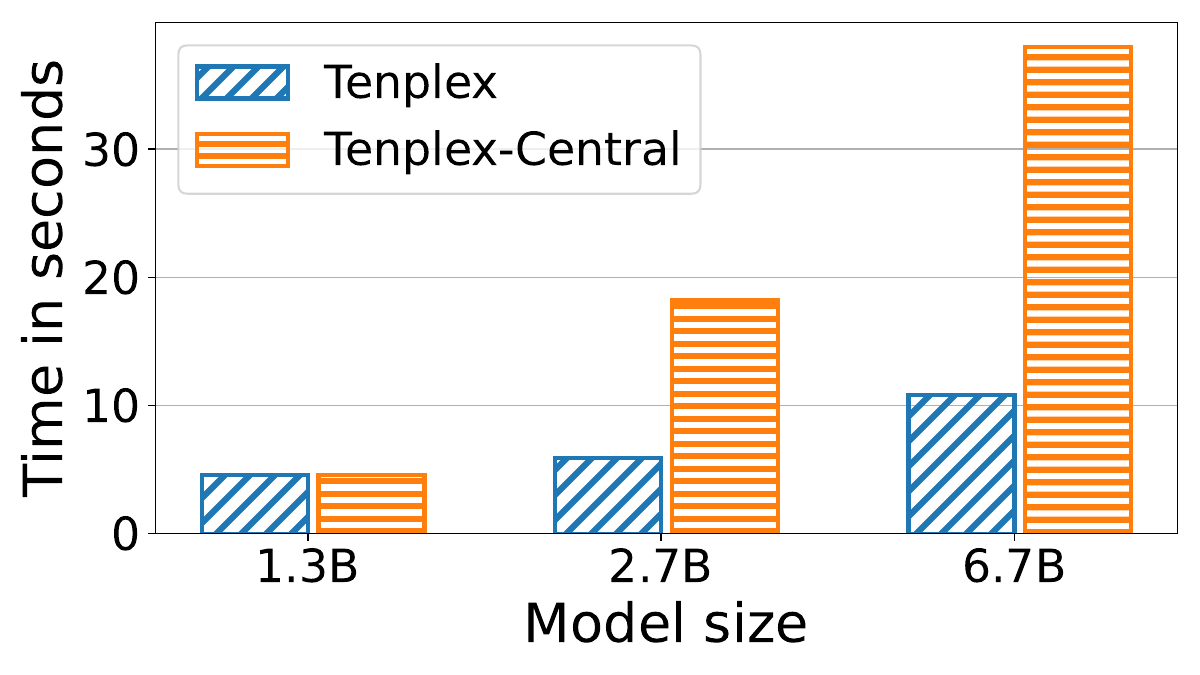}
    \captionsetup{width=.9\linewidth, aboveskip=1pt, belowskip=1pt}
    \caption{Pipeline parallelism}
    \label{fig:state-transformer-time-size-pp}
  \end{subfigure}
  \hfill
  \begin{subfigure}[t]{0.33\textwidth}
    \includegraphics[width=.9\columnwidth]{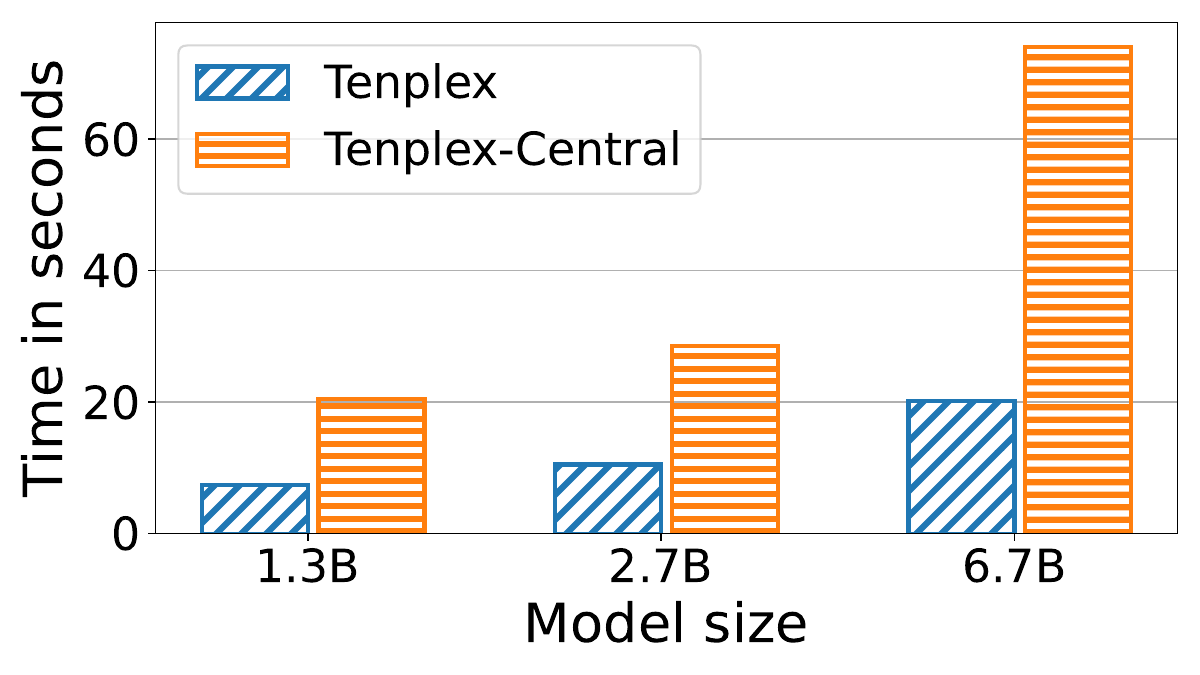}
    \captionsetup{width=.9\linewidth, aboveskip=1pt, belowskip=1pt}
    \caption{Tensor parallelism}
    \label{fig:state-transformer-time-size-mp}
  \end{subfigure}
  \caption{Reconfiguration time with different parallelizations}
  \label{fig:state-transformer-time-size}
\end{figure*}

\begin{figure*}[t]
  \centering
  \begin{subfigure}[t]{0.33\textwidth}
    \includegraphics[width=.9\columnwidth]{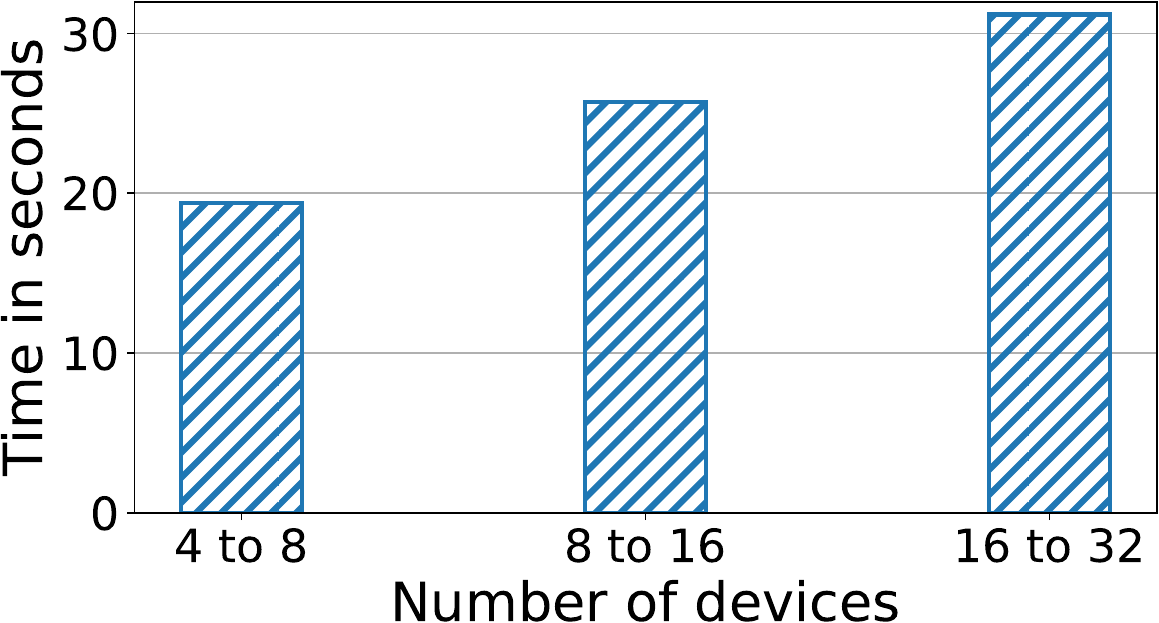}
    \captionsetup{width=.9\linewidth, aboveskip=1pt, belowskip=1pt}
    \caption{Data parallelism}
    \label{fig:state-transformer-nd-dp}
  \end{subfigure}
  \hfill
  \begin{subfigure}[t]{0.33\textwidth}
    \includegraphics[width=.9\columnwidth]{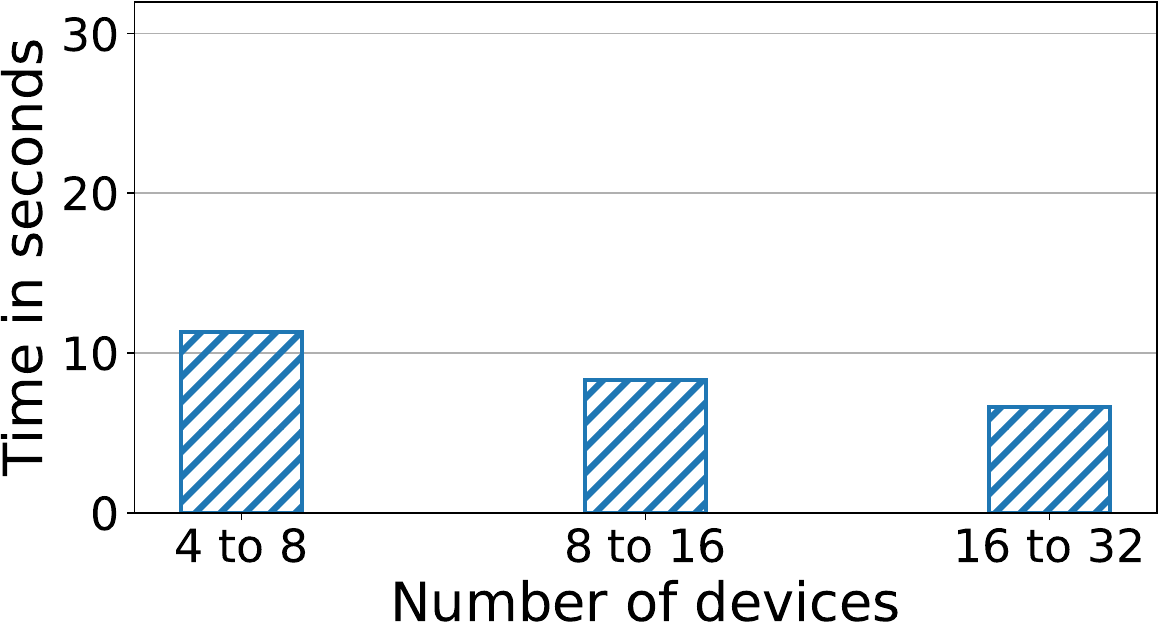}
    \captionsetup{width=.9\linewidth, aboveskip=1pt, belowskip=1pt}
    \caption{Pipeline parallelism}
    \label{fig:state-transformer-nd-pp}
  \end{subfigure}
  \hfill
  \begin{subfigure}[t]{0.33\textwidth}
    \includegraphics[width=.9\columnwidth]{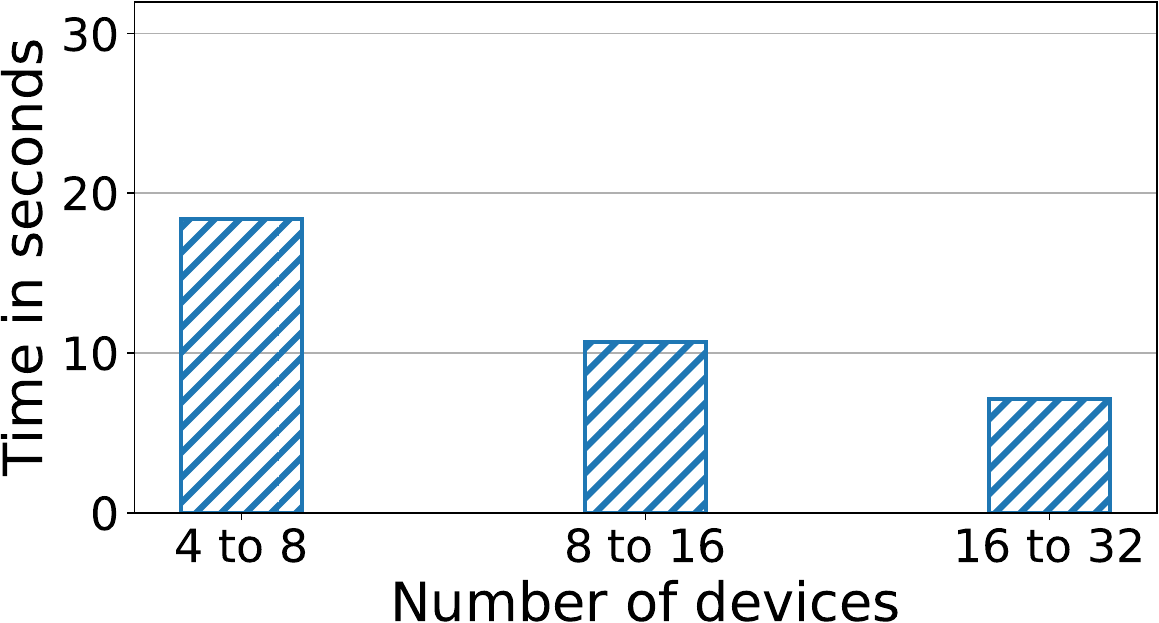}
    \captionsetup{width=.9\linewidth, aboveskip=1pt, belowskip=1pt}
    \caption{Tensor parallelism}
    \label{fig:state-transformer-nd-mp}
  \end{subfigure}
  \caption{Reconfiguration time with different cluster sizes}
  \label{fig:state-transformer-time-parallelism}
\end{figure*}

\subsection{Reconfiguration overhead}
\label{sec:eval:diff-reconfiguration}

This experiment compares the reconfiguration approach of \sys with (i)~a model library of an elastic DL system~(\textsf{DeepSpeed}) and (ii)~a virtual device approach that performs full GPU state migration~(\textsf{Singularity}).

We use the GPT-3 XL model with the Wikipedia dataset on the on-premise cluster. We perform one experiment that scales down resources from 16~to 8~GPUs and another that scales up from 8 to 16~GPUs. Since Singularity~\cite{singularity} is a closed-source system, we report numbers from a similar experiment in its paper, run on similar hardware.

\F\ref{fig:comparison-scaling-latency} shows the reconfiguration time. When changing from 8 to 16~GPUs, \sys requires 24\% less time than DeepSpeed and 10\% less time than Singularity. Singularity is slower, because, besides the training state, it also moves the full GPU device state. DeepSpeed suffers from the fact that it does not include an explicit mechanism for notifying DeepSpeed about reconfiguration but instead uses its failure detection mechanism. The state management approach of \sys is the fastest because it minimizes state movement due to its awareness of data locality.

The difference becomes larger when scaling from 16 to 8~GPUs: \sys needs 64\% less time than DeepSpeed and 43\% less than Singularity. In this case, DeepSpeed relies on Torch Distributed Elastic's failure mechanism, which increases time; Singularity must copy the full GPU state even though there already is a model replica on the GPUs.

We also compare \sys{}'s overhead to \textsf{Horovod}, a distributed training library without elasticity support, and \textsf{Horo\-vod-Elastic}, which also supports scaling under data parallelism only by periodically checkpointing the model state. We deploy a ResNet50 model with the ImageNet dataset in the on-premise cluster and measure throughput when training on 2~GPUs.

\F\ref{fig:throughput} compares the training throughput, measured as samples per second, for \textsf{Horovod}, \textsf{Elastic Horovod}, and \textsf{\sys}. \textsf{Horovod} achieves 438\unit{images/s}, but with elasticity support, the throughput drop to 418\unit{images/s}. \sys achieves a throughput of 430\unit{images/s}, which is about the same as regular \textsf{Horovod} without elasticity support.

We conclude that, despite supporting dynamic reconfiguration for multi-dimensional parallelism, \sys matches \textsf{Horovod}{}'s performance. \sys{}'s state management does not interfere with training, and reconfiguration only incurs a cost when resources change. Unlike \textsf{Horovod Elastic}, \sys avoids explicit state synchronization in a blocking manner after a user-defined number of steps, which interrupts training progress.

\subsection{Impact of parallelization type}
\label{sec:eval:impact-model-size}

Next, we examine the impact of the parallelization configuration on reconfiguration time for different model sizes. We deploy \textsf{\sys} and \textsf{\sys-Central}, which manages the state in a single node, with the different GPT-3 models on the on-premise cluster. For data parallelism~(D), we change the configuration from $(T,P,D) = (4,2,1)$ to $(4,2,2)$; for pipeline parallelism~(P) from $(4,2,1)$ to $(4,4,1)$; and for tensor parallelism~(T) from $(4,2,1)$ to $(8,2,1)$.

\F\ref{fig:state-transformer-time-size} shows the reconfiguration time for the different parallelization configurations and model sizes. With data parallelism (\F\ref{fig:state-transformer-time-size-dp}), \textsf{\sys-Central} with GPT-3 6.7\unit{B} takes 4$\times$ longer than \sys, because of the limited network bandwidth of a single worker in comparison with a distributed peer-to-peer state reconfiguration.
We observe similar behavior for pipeline and tensor parallelism: under pipeline parallelism (\F\ref{fig:state-transformer-time-size-pp}), \textsf{\sys-Central} takes 3.5$\times$ longer and, under tensor parallelism (\F\ref{fig:state-transformer-time-size-mp}), it takes 3.7$\times$ longer.
The only exception is pipeline parallelism with 1.3\unit{B} parameters. In this case, network bandwidth does not become a bottleneck, because the parallelization configuration does not involve splitting and merging sub-tensors.

We conclude that centralized state management becomes a bottleneck with many model parameters, due to the limited network bandwidth and the reduced parallelism when all state transformations are performed by one worker.

\begin{figure*}[t]
  \centering
  \begin{subfigure}[t]{0.33\textwidth}
    \centering
    \includegraphics[width=.9\columnwidth]{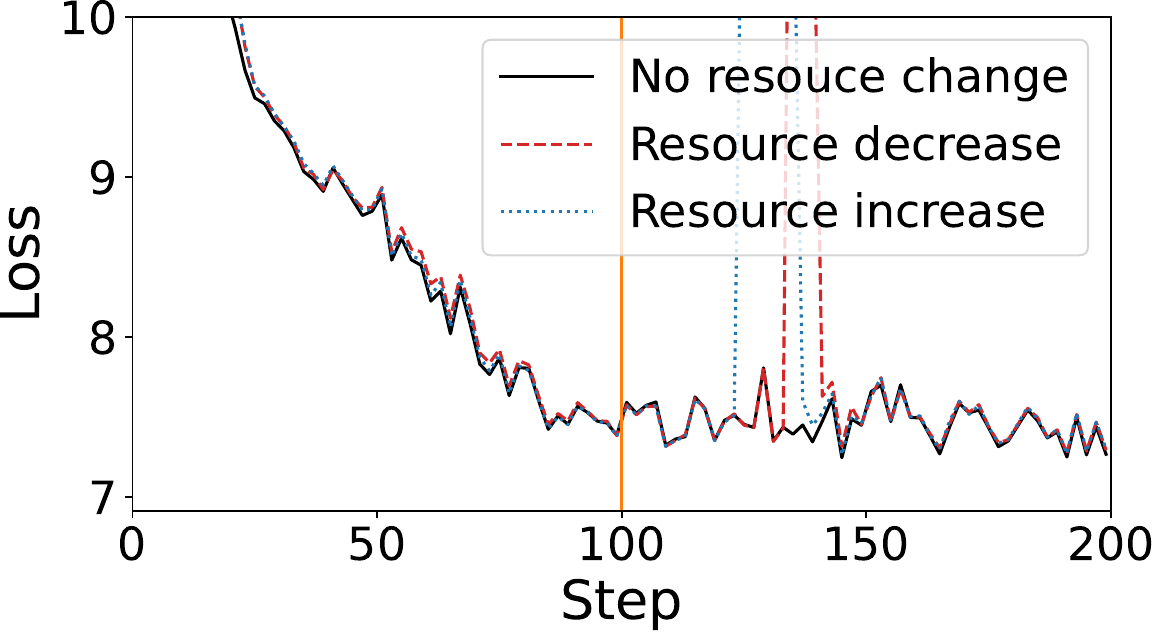}
    \captionsetup{width=.9\linewidth, aboveskip=1pt, belowskip=1pt}
    \caption{Data parallelism}
    \label{fig:scale-dp}
  \end{subfigure}
  \hfill
  \begin{subfigure}[t]{0.33\textwidth}
    \includegraphics[width=.9\columnwidth]{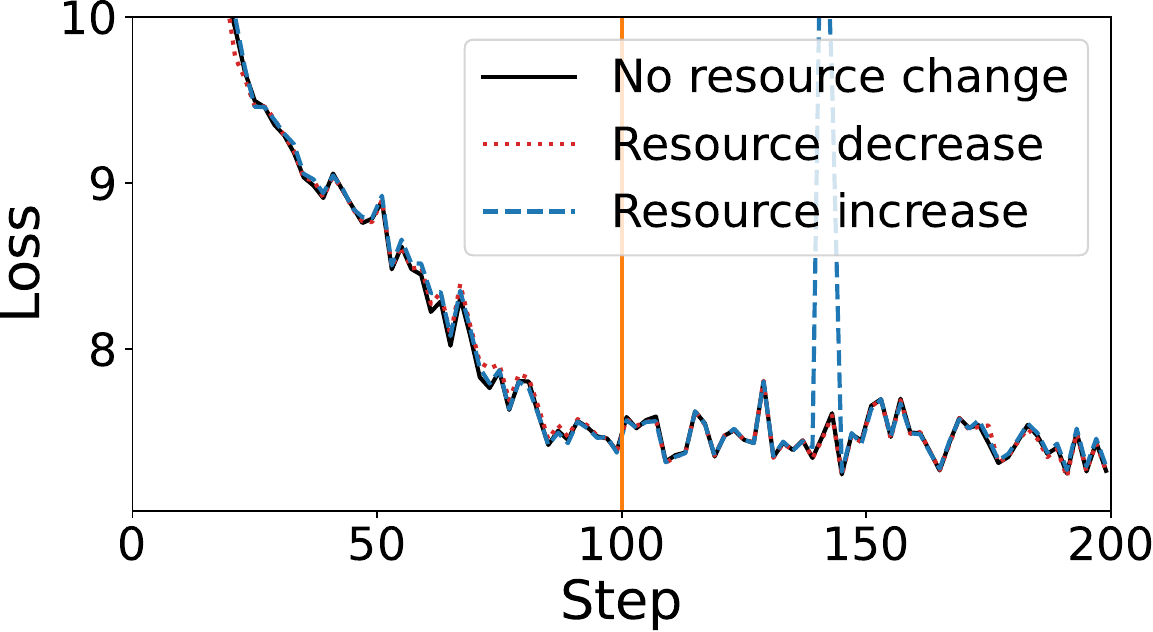}
    \captionsetup{width=.9\linewidth, aboveskip=1pt, belowskip=1pt}
    \caption{Pipeline parallelism}
    \label{fig:scale-pp}
  \end{subfigure}
  \hfill
  \begin{subfigure}[t]{0.33\textwidth}
    \centering
    \includegraphics[width=.9\columnwidth]{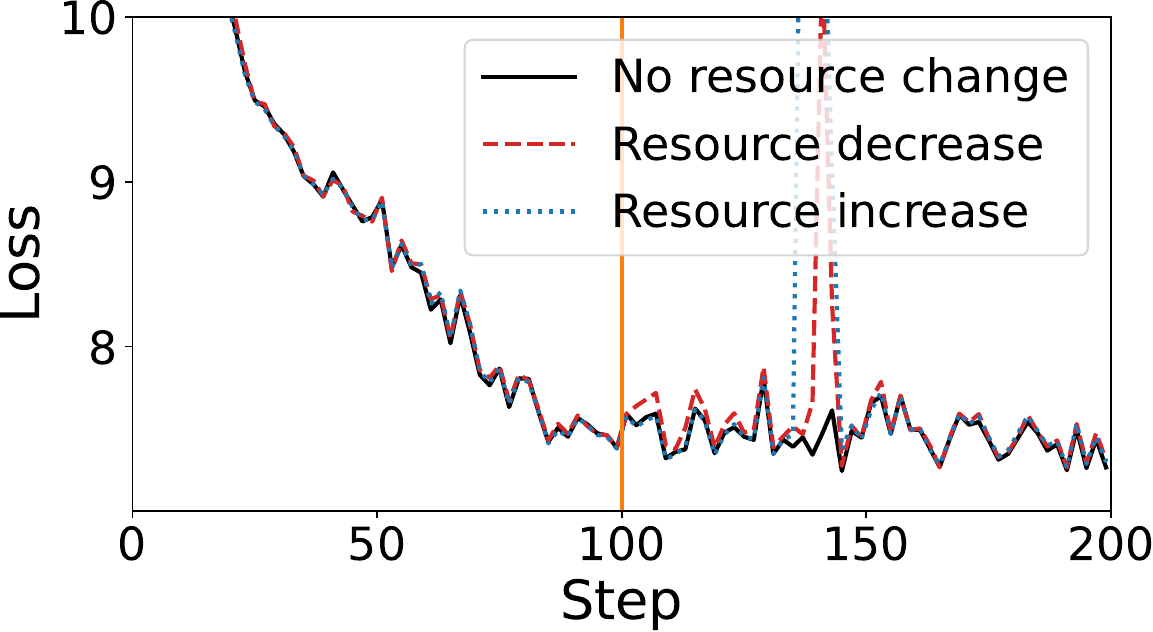}
    \captionsetup{width=.9\linewidth, aboveskip=1pt, belowskip=1pt}
    \caption{Tensor parallelism}
    \label{fig:scale-mp}
  \end{subfigure}
  \caption{Model convergence with reconfiguration}
  \label{fig:scaling-mdp}
\end{figure*}

\subsection{Impact of cluster size}
\label{sec:eval:impact-resource-count}

We want to explore how \sys is affected by the GPU cluster size. In this experiment, we keep the model size fixed but change the GPU resources in the cluster to evaluate how the cluster size and parallelization configuration impact reconfiguration time.

We use the GPT-3 XL on the Wikipedia dataset deployed in the 32-GPU cloud testbed. We scale the resources from 4 to 8, 8 to 16, and 16 to 32~GPUs for data, tensor, and pipeline parallelism, respectively. For each parallelization configuration, if the number of GPUs doubles, the degree of parallelism also doubles. We compare \sys with the baseline \textsf{\sys-Central}, as it is the only baseline that supports full multi-dimensional parallelism.

\F\ref{fig:state-transformer-time-parallelism} shows the reconfiguration time with different device counts. For data parallelism~(\F\ref{fig:state-transformer-nd-dp}), the time increases linearly with the number of GPUs, because the number of model replicas is proportional to the parallelism degree; for pipeline parallelism~(\F\ref{fig:state-transformer-nd-pp}), the reconfiguration time decreases with the number of devices, because the model size is constant and the total network bandwidth increases with the GPU count; for tensor parallelism~(\F\ref{fig:state-transformer-nd-mp}), the time decreases with the GPU count, because the model size is constant and the network bandwidth increases with the devices.

Comparing data, tensor, and pipeline parallelism, the reconfiguration time is the highest with data parallelism, because the amount of data increases with the number of replicas. While the amount of data stays constant with pipeline and tensor parallelism, tensor parallelism must split and merge sub-tensors. The reconfiguration time is lowest with pipeline parallelism, which only needs repartitioning.

\subsection{Impact on model convergence}
\label{sec:eval:convergence}

Finally, we evaluate \sys{}'s impact on model convergence. For this, we use the BERT-large model with the OpenWebText dataset deployed on the on-premise cluster. At training step~100, we either increase or decrease the resources and compare them to a baseline without change.

\F\ref{fig:scaling-mdp} shows the model convergence as the loss over the training steps. With data parallelism~(\F\ref{fig:scale-dp}), \ie changing the parallelization configuration from $(1,1,4)$ to $(1,1,8)$, the loss does not diverge when the resources increase/decrease because \sys maintains consistent hyper-parameters and consistent data order. With pipeline parallelism~(\F\ref{fig:scale-pp}), \ie changing from $(1,4,1)$ and $(1,8,1)$, and with tensor parallelism~(\F\ref{fig:scale-mp}), \ie changing from $(4,1,1)$ and $(8,1,1)$, convergence is equally unaffected.




\section{Related Work}
\label{sec:related_work}


\myparr{Elastic ML systems}~\cite{kungfu, horovod} support resource changes but typically only adapt data parallelism by adding/removing model replicas to/from GPU workers---they leave more general state management for multi-dimensional parallelism to users, which hinders adoption. Torch Distributed Elastic~(TDE)~\cite{pytorch_elastic} only supports changes to data parallelism because it cannot re-partition state during training. Varuna~\cite{varuna} needs the user to define partitioning points for pipeline parallelism. It then only reconfigures data/pipeline parallelism, which prevents it from supporting strategies with tensor or sequence parallelism. Spotnik~\cite{spotnik2020} supports elasticity over transient GPU resources but only supports data parallelism. Hydrozoa~\cite{Hydrozoa} supports static \mdp before training but only allows for dynamic changes to data parallelism. GoldMiner~\cite{zhao2023goldminer} focuses on adapting data preprocessing in ML systems, but it does not support multi-dimensional parallelism. In contrast, \sys has a generic state abstraction for multi-dimensional parallelism and leverages it for automatic reconfiguration.

In terms of reconfiguration performance, TDE and Varuna follow a centralized approach to state management, \ie they load/save state to one machine. This can result in a performance bottleneck when re-configuring a large model state.


Dynamic GPU scheduling systems, \eg Lyra~\cite{li2023lyra}, adjust the numbers of GPUs assigned to jobs. They rely, however, on elastic ML libraries for the reconfiguration of jobs, making them complementary to \sys.

\myparr{Checkpointing systems} store snapshots of model parameters. When resources change, \eg after failure, the DL system retrieves the latest checkpoint before the failure and resumes training. CheckFreq~\cite{checkfreq} dynamically adjusts the checkpointing frequency, while Check-N-Run~\cite{checknrun} uses lossy compression, trading accuracy for storage efficiency. ServerlessLLM~\cite{fu2024serverlessllm} provides fast checkpoint loading in a serverless GPU cluster, but it does not support checkpoint reconfiguration as \sys does. Gemini~\cite{DBLP:conf/sosp/WangJZZFNW23} and Oobleck~\cite{DBLP:conf/sosp/JangYZJC23} provide fast checkpointing: while the former stores checkpoints in CPU memory, the latter maintains checkpoints in GPU memory. These approaches focus only on the performance of failure recovery, as opposed to the generic runtime reconfiguration of \sys. 

\mypar{DL job parallelization} DL systems have built-in parallelization support: PyTorch~\cite{pytorch} and TensorFlow~\cite{tensorflow} offer data parallelism; JAX~\cite{jax} has \emph{pmap} and \emph{xmap} functions to parallelize computation; and MindSpore~\cite{mindspore} uses \emph{auto parallel} search to find an effective parallelization strategy. All of these approaches only support a subset of multi-dimensional parallelism and do not offer runtime reconfiguration.

Unity~\cite{unity} and Alpa~\cite{alpa} search for an optimal distribution plan and implement a suitable runtime for it. \sys uses such parallelizers when making reconfiguration decisions.

Google XLA~\cite{sabne2020xla} and MindSpore have resharding operators, which are needed for automatic parallelism. These operators, however, are applied to a single GPU instead of a GPU cluster, and therefore cannot handle the reconfiguration of distributed DL jobs with multi-dimensional parallelism, as supported by \sys.

\myparr{Distributed data analytics} systems may have support for elasticity: some systems, MapReduce~\cite{mapreduce} or stream processing~\cite{spark}, can change the resources of jobs. In contrast, DL training jobs with \mdp require the concept of a tensor to split the model parameters and datasets correctly for strategies with tensor parallelism. Elastic Memory~\cite{elasticmemory} supports some DL jobs with data parallelism but does not handle other parallelism strategies. Cruise~\cite{cruise} lacks abstractions for tensor splitting.

\section{Conclusion}

We described \sys, a dynamic state management library for DL jobs with multi-dimensional parallelism. By describing the state as a parallelizable tensor collection~(PTC), \sys  generates efficient reconfiguration plans when the underlying GPU resources
 for the job 
 change at runtime. Its distributed state transformers implement the reconfiguration plan on each GPU with a minimum amount of data movement between workers. Therefore, \sys is a step towards making large-scale long-running deep learning jobs fully adaptive to resource changes.

\bibliographystyle{ACM-Reference-Format}
\bibliography{\jobname}

\end{document}